\documentclass[10pt,journal,compsoc]{IEEEtran}

\usepackage[nocompress]{cite}

\usepackage[utf8]{inputenc}
\usepackage[T1]{fontenc}
\usepackage[hidelinks]{hyperref}
\usepackage{tabulary}
\usepackage{booktabs}
\usepackage{amsmath}
\usepackage{amsfonts}
\usepackage{amssymb}
\usepackage{graphicx}
\usepackage{xspace}
\usepackage{multirow}
\usepackage{array}
\usepackage{url}
\usepackage{ragged2e}
\usepackage{caption}
\usepackage{syntax}
\usepackage{mdframed}
\usepackage{subcaption}
\usepackage{balance}
\usepackage[linesnumbered,ruled,onelanguage,english]{algorithm2e}

\newcommand\MS{\mathit{MS}}
\newcommand\DM{\mathit{DM}}
\newcommand\EM{\mathit{EM}}

\begin{document}

\title{Sentinel: A Hyper-Heuristic for the \\Generation of Mutant Reduction Strategies}

\author{Giovani~Guizzo,
	Federica~Sarro,
	Jens~Krinke,
	and~Silvia~R.~Vergilio%
	\IEEEcompsocitemizethanks{
		\IEEEcompsocthanksitem G. Guizzo, F. Sarro (corresponding author) and J. Krinke are with the University College London, London WC1E 6BT, United Kingdom.\protect \\
		E-mail: \{g.guizzo, f.sarro, j.krinke\}@ucl.ac.uk.
		\IEEEcompsocthanksitem S. R. Vergilio is with the Department of Informatics, Federal University of Paraná, Curitiba, PR, Brazil.\protect \\
		E-mail: silvia@inf.ufpr.br}%
	\thanks{Manuscript received September 14, 2018; revised September 26, 2018.}}

\markboth{IEEE Transactions on Software Engineering}{Guizzo \MakeLowercase{\textit{et al.}}: Sentinel}

\IEEEtitleabstractindextext{
\begin{abstract}
Mutation testing is an effective approach to evaluate and strengthen software test suites, but its adoption is currently limited by the mutants' execution computational cost. Several strategies have been proposed to reduce this cost (a.k.a. mutation cost reduction strategies), however none of them has proven to be effective for all scenarios since they often need an ad-hoc manual selection and configuration depending on the software under test (SUT).

In this paper, we propose a novel multi-objective evolutionary hyper-heuristic approach, dubbed Sentinel, to automate the generation of optimal cost reduction strategies for every new SUT. We evaluate Sentinel by carrying out a thorough empirical study involving 40 releases of 10 open-source real-world software systems and both baseline and state-of-the-art strategies as a benchmark. We execute a total of 4,800 experiments, and evaluate their results with both quality indicators and statistical significance tests, following the most recent best practice in the  literature.

The results show that strategies generated by Sentinel outperform the baseline strategies in 95\% of the cases always with large effect sizes. They also obtain statistically significantly better results than state-of-the-art strategies in 88\% of the cases, with large effect sizes for 95\% of them. Also, our study reveals that the mutation strategies generated by Sentinel for a given software version can be used without any loss in quality for subsequently developed versions in 95\% of the cases.

These results show that Sentinel is able to automatically generate mutation strategies that reduce mutation testing cost without affecting its testing effectiveness (i.e.\ mutation score), thus taking off from the tester's shoulders the burden of manually selecting and configuring strategies for each SUT.
\end{abstract}
	\begin{IEEEkeywords}
		Mutation Testing, Mutant Reduction, Software Testing, Grammatical Evolution, Hyper-Heuristic, Search Based Software Testing, Search Based Software Engineering
	\end{IEEEkeywords}
}

\maketitle

\IEEEdisplaynontitleabstractindextext

\IEEEpeerreviewmaketitle

\IEEEraisesectionheading{\section{Introduction}\label{sec:Introduction}}

\IEEEPARstart{M}{utation} testing is a white-box testing technique used to evaluate the capability of a  test suite in revealing faults. Even though this technique is efficacious in evaluating and guiding the creation of good test cases, its main disadvantage is the great computational cost related to the execution of mutants~\cite{Jia2011,Papadakis2017}. Depending on the number of mutants and test cases, the creation of new test cases, evaluation of equivalent mutants, and the eventual execution of alive mutants might be very expensive, both in terms of machine computational power and human resources.

This problem can be diminished by reducing the number of mutants, and consequently reducing the mutant execution costs~\cite{Pizzoleto2019}. This can be done using mutant reduction strategies~\cite{Jia2011,Papadakis2017,Pizzoleto2019}, however, experiments reported in the literature show that no strategy is the best for all scenarios~\cite{Pizzoleto2019}. Moreover, even the simplest strategies have some parameters to be configured depending on the software under test (SUT). Given the great number of available mutant reduction strategies and the tuning they need, a human tester may find it difficult to choose and manually configure a strategy for their SUT. Let us consider the following scenario as an example: in continuous integration practice, all developers' working copies are merged to a shared mainline repository multiple times a day, and it is very common that the engineers test the software each time they commit their code to this shared repository~\cite{AlshahwanCH0MMM19}. Similarly, an open-source software under development must be rigorously tested before each commit because there are multiple contributors to the shared code. In such scenarios, performing mutation testing using all generated mutants is likely to be infeasible and it is definitively detrimental if the testers have to wait for the mutation testing to finish in order to develop further code and/or test cases. Certainly, the mutants' execution cost can be reduced by using more powerful machines and parallelism, yet the time that could be saved by using mutant reduction strategies is meaningful for nowadays programs with tens of thousands of mutants and is beneficial for the environment~\cite{Calero2015}. Moreover, an arbitrarily selected and configured strategy can result in unsatisfactory cost reduction, test effectiveness maintenance or, even worse, both. In the long run, such arbitrarily chosen strategy could end up not saving as much time as a strategy tailored for a specific testing scenario. Determining the best strategy and its best configuration for a given scenario and context usually is only possible through an experimental evaluation, which can be seen as an optimisation problem itself~\cite{Eiben2011}.

In this sense, we advocate that the usage of hyper-heuristics~\cite{Burke2010} is a viable option to automatically generate and select mutant reduction strategies. Those strategies can be optimised to reduce the mutation execution cost while maintaining the mutation score, and the hyper-heuristic itself can relieve the tester from the burden of selection and configuration tasks. To this end, we introduce Sentinel, an off-line multi-objective learning hyper-heuristic based on Grammatical Evolution (GE)~\cite{Ryan1998} to automatically generate mutant reduction strategies (low-level heuristics) for the mutation testing of Java programs. To the best of our knowledge, there is no previous work that investigates the automatic generation, nor the selection and configuration of mutant reduction strategies.

To evaluate Sentinel, we first compare it with a Random Hyper-Heuristic as a sanity check. Then, we compare the strategies generated by Sentinel with three state-of-the-art conventional strategies: Random Mutant Sampling, Random Operator Selection and Selective Mutation. This analysis has been carried out using 10 real-world open-source systems with four different versions each (for a total of 40 releases and 4,800 comparisons).

Our empirical results show that Sentinel is able to generate strategies providing statistically significantly better results than a Random Hyper-Heuristic for all the systems always with large effect size. Furthermore, Sentinel generates strategies that outperform conventional strategies proposed in previous work. For 70 out of 80 comparisons (i.e.\ 88\%), Sentinel outperforms these state-of-the-art strategies with statistically significant differences and obtains favourable large effect sizes for 227 out of 240 cases (i.e.\ 95\%).

To summarise, the main contributions of this work are:

\begin{itemize}
	\item[--] the proposal of Sentinel, a GE based multi-objective hyper-heuristic to automatically generate mutant reduction strategies that are able to provide the best trade-off between mutation execution time reduction and maintenance of the global mutation score;
	\item[--] an empirical study investigating 10 real-world open source systems (for a total of 40 releases), which is the largest done so far for investigating hyper-heuristic for mutation testing;
	\item[--] a public repository with the data used in this work, which allows for replication and extension of our study and can be also used for other studies on mutation testing~\cite{SentinelWebsite};
	\item[--] an open-source reference implementation of Sentinel in Java to allow its usage and extension~\cite{SentinelGH}.
\end{itemize}

The next sections present a background on the main topics of this work (Section~\ref{sec:Background}), the description of Sentinel (Section~\ref{sec:Sentinel}) and its empirical evaluation (Sections~\ref{sec:Empirical Study} and~\ref{sec:Empirical Study:Results}), related work (Section~\ref{sec:Related Work}), and final remarks (Section~\ref{sec:Conclusion}).

\section{Background}
\label{sec:Background}

\subsection{Mutation Testing}
\label{sec:Mutation Testing}

Mutation testing is based on the concept of mutants, where a mutant is a derivation of the original software program with a small syntactic change. Each kind of syntactic change is introduced by using a mutation operator. A test case is said to kill a mutant if the test case yields a different output when executing the original and the mutated program. If a mutant is killed by a test case, it means that the test will be able to reveal this fault. If a mutant is still alive, then the tester may use this information to create stronger test cases able to kill such a mutant.

Usually, the testing criterion for assessing the quality of a test set is the mutation score, which computes the number of non-equivalent mutants that are killed by the test set. A mutant is called equivalent if it yields the same output of the program under test. For such a mutant, there is no test data capable to distinguish it from the original program. Equation~\ref{eq:Mutation Score} shows how the mutation score is computed: \begin{equation}
\MS(T,M) = \dfrac{\DM(T,M)}{|M| - \EM(M)}
\label{eq:Mutation Score}
\end{equation}

\noindent where $M$ is a set of mutants; $T$ is a set of test cases; $\MS(T,M)$ is the mutation score obtained when executing $T$ against $M$; $\DM(T,M)$ is the number of mutants in $M$ killed by $T$; $|M|$ is the number of mutants in $M$; and $\EM(M)$ is the number of equivalent mutants in $M$.

The greater the mutation score, the better the test set is in revealing the faults represented by the mutants. A test set is called ``adequate'' when it is able to kill all the non-equivalent mutants, i.e.\ when it obtains a mutation score of $1.0$ (100\%). Ideally, a test set should kill all mutants of a program, but that is a tricky task. It is very costly (regarding both computational cost and human effort) to create test cases to kill mutants and execute all alive mutants every time a test case is created. Moreover, determining $\EM(M)$ is (generally) undecidable, and the mutation score is usually computed assuming that all mutants are non-equivalent.

\subsection{Mutant Reduction}
\label{sec:Mutant Reduction}

Given the great cost of mutation testing, one can try to reduce its cost by using mutation cost reduction strategies. These strategies can be classified into: ``do fewer'', ``do faster'' and ``do smarter''~\cite{Offutt2001,Pizzoleto2019}. ``Do faster'' strategies consist in the faster execution of mutants such as running compiled programs instead of interpreted programs, mutant schema, parallelism, and others. ``Do smarter'' strategies try to avoid the full execution of mutants, such as Weak Mutation strategies~\cite{Howden1982} that evaluate the state of the mutant right after executing the faulty instruction. ``Do fewer'' strategies try to reduce the set of mutants that need to be executed. In this work we focus on ``do fewer'' strategies.

When reducing mutant sets, the tester must assert that the effectiveness of the reduced mutant set is not compromised, otherwise the reduced set will not be sufficient to guide the selection/creation of good test cases. The mutant reduction problem can be defined as the search for a subset of mutants $M'$ derived from all mutants $M$, such that: i)~the approximation $\MS(T', M) \approx \MS(T, M)$ (herein called ``approaching'') holds when selecting a subset of test cases $T'$ from the SUT's existing test set $T$; or ii)~$\MS(T'', M)$ is maximised when creating a new test set $T''$. Either way, it is the subset of mutants $M'$ that is used to guide the attainment of test cases. If the test effectiveness is not compromised during reduction, then only $|M'|$ mutants are needed to find an adequate set of test cases $T'$ or $T''$ that can efficiently kill all killable mutants in $M$.

Our notion of test effectiveness maintenance during test case selection is similar to the test effectiveness maintenance presented by Amman et al.~\cite{Ammann2014}:\begin{quote}
	``Formally, a subset of $T$, denoted $T_{\mathit{maintain}}$, maintains the mutation score with respect to $M$ (and $T$) if for every mutant $m$ in $M$, if $T$ kills $m$ then $T_{\mathit{maintain}}$ kills $m$.''
\end{quote}

This is analogous to our approaching definition $\MS(T', M) \approx \MS(T, M)$, because if $\MS(T, M) = \MS(T', M)$ then we can also state that $T'$ is $T_{\mathit{maintain}}$ according to their definition. Notice that, in the context of our work, $T$ is the complete test suite of the SUT and represents the pool of all available test cases for that SUT. It serves as a ``ground truth'' of what the subsets $T'$ and $M'$ can achieve in terms of test effectiveness. Therefore, the mutation score $\MS(T', M)$ cannot overshoot the global mutation score $\MS(T, M)$, since $T'$ is a subset of $T$. For this reason, in this work, we use the term ``approaching'' instead of ``approximating'' to describe $\MS(T', M) \approx \MS(T, M)$. Moreover, $T$ does not usually contain all possible test cases for that SUT, however, $T$ can serve as a baseline to guide the mutant reduction without the need of generating new tests or evaluating mutants to determine their equivalence.

We can find several strategies in the literature that are used to reduce the mutant set~\cite{Pizzoleto2019}. These strategies (herein called conventional) can sample a subset of mutants~\cite{Mathur1994,Wong1995,Sahinoglu1990}, apply fewer mutation operators~\cite{Offutt1993,Mathur1991,Delamaro2014}, find a set of essential mutation operators to reuse~\cite{Barbosa2001,Offutt1996,Vincenzi1999,Namin2008}, group mutants~\cite{Hussain2008,Ji2009}, or perform higher order mutation~\cite{Harman2010}. More recent strategies are based on search based algorithms~\cite{Silva2016}. Such strategies are part of the Search Based Software Engineering (SBSE) field~\cite{Harman2012}, which in turn aims at solving hard software engineering problems with the aid of search based algorithms.

The most common conventional strategies were identified from the literature~\cite{Pizzoleto2019,Jia2011,Papadakis2017} and used in the experiments of this work. These strategies are explained next.

Random Mutant Sampling (RMS)~\cite{Mathur1994,Wong1995,Sahinoglu1990} randomly selects a set of mutants from the pool of all mutants generated for a given system. With this strategy, the tester selects a subset of mutants before executing them, thus the mutants execution is supposedly faster than executing all mutants. This strategy needs to be configured with a percentage of mutants to be sampled. Usually, the more mutants selected, the better the effectiveness of the sampled mutant set, but the greater the time spent with the mutation testing.

Random Operator Selection (ROS)~\cite{Offutt1993,Mathur1991,Delamaro2014} randomly selects a subset of mutation operators to generate mutants. Instead of generating and then selecting mutants like RMS, with ROS the selection is done before any mutant is generated. The advantage of this strategy is that, not only fewer mutants are executed, but also fewer operators are applied. This strategy requires one parameter that can be a fixed number or a percentage of operators to execute. Usually, the more operators executed, the more mutants generated, but the costlier the mutation activity.
	
Selective Mutation (SM)~\cite{Offutt1993,Mathur1991,Delamaro2014,Barbosa2001,Offutt1996,Vincenzi1999,Namin2008} is used to select a subset of operators. ROS is similar to SM but while ROS selects operators at random, SM uses information about the mutation testing activity in order to guide the selection of operators. In this work we used the SM version that discards the $n$ operators that generate the most mutants. This can potentially discard operators that generate a large number of redundant mutants which can be killed by a single test case. However, while it is trivial to assess which operators generate the most mutants, it is not a trivial task to assess how many of those operators should be excluded from the mutation testing activity, thus the tuning of $n$ should be considered beforehand.

\subsection{Multi-Objective Evolutionary Algorithms}
\label{sec:Sentinel:Multi-Objective Optimisation}

Evolutionary Algorithm (EA) are based on the theory of evolution~\cite{Eiben2003}, in which a population evolves over several generations by means of natural selection and reproduction. In EAs, a solution is an individual in the population represented by a chromosome (genotype). Usually the representation is an array of genes (e.g.\ bit, float or integer) on which the algorithm performs perturbations to find different solutions. Over several generations, parent individuals are selected to reproduce (crossover) and generate offspring that carry their genes. Then, the offspring are mutated for diversity and the best solutions survive to the next generation. In each generation, the solutions are evaluated using fitness functions, which compute their overall quality. This continues until a stopping condition is met.

Mono-objective EAs are straightforward: the solutions are given a fitness value each, ranked accordingly, and then the ones with the best fitness are selected to reproduce and survive during the evolutionary process. At the end of the optimisation, the engineer selects the solution with the best fitness. On the other hand, Multi-Objective Evolutionary Algorithms (MOEAs)~\cite{Coello2007} optimise simultaneously two or more objectives with equal weight considering the Pareto dominance concept~\cite{Coello2007}. If all objectives $z \in Z$ of a problem are of minimisation, a solution $x$ is said to dominate a solution $y$ ($x \prec y$) if it is better or equal in all objectives and better in at least one: \begin{equation}
\begin{split}
\forall z \in Z: z(x) \leq z(y)\\
\exists z \in Z: z(x) < z(y)
\end{split}
\end{equation}

\noindent otherwise the solutions are called non-dominated. Non-dominated solutions are not easy to compare, because each one is a viable solution for the problem at hand and represents a satisfactory trade-off between objectives. It is up to the engineers to select the solution that best fits their needs.

When performing experiments and comparing the results of algorithms however, each resulting Pareto front must be evaluated as a whole so that the general performance of the algorithm can be assessed. Quality indicators~\cite{Zitzler2003} are usually applicable to quantify the quality of the results. The main goal of quality indicators is to provide a means of computing a single value (unary indicators) that represents the quality of a Pareto front, or quantifying the difference between two fronts (binary indicators). In this work we use Hypervolume (HV) and Inverted Generational Distance (IGD)~\cite{Zitzler2003}, two of the most common unary indicators in the literature also used in SBSE~\cite{Guizzo2015,Guizzo2016,Guizzo2017,Guizzo2017b,FerrucciHRS13,ZhangHJS15, SarroPH16,Sarro2017}.

HV~\cite{Zitzler2003} computes the area or volume of the objective space that is dominated by a given front in relation to a reference point (usually the worst possible point). The greater the HV value, the greater the area of the objective space a front dominates, thus the better the algorithm in terms of finding non-dominated solutions. IGD~\cite{Zitzler2003} computes the distance between the true Pareto front (the best solutions found so far) and a given front. This is done by summing the distance in the objective space from each solution of the true Pareto front to the nearest solution of the front being evaluated. Therefore, the lower the IGD value, the closer the front is to the true Pareto front, and consequently the closer the solutions are to the best found.

\subsection{Hyper-Heuristics}
\label{sec:Hyper-Heuristics}

Hyper-Heuristics are used to select or generate the best low-level heuristics instead of trying to solve a problem directly~\cite{Burke2010}. Hence, instead of acting over the search space, hyper-heuristics search the heuristic space for good heuristics that can solve the problem.

Hyper-Heuristics can be of on-line or off-line nature. On-line hyper-heuristics select or generate the low-level heuristics during the optimisation, whereas off-line hyper-heuristics apply the training before the optimisation and then reuse the selected or generated heuristics in unseen instances of the problem. A low-level heuristic is any heuristic that is being selected or generated by a hyper-heuristic, being either a construction or a perturbation heuristic. Construction low-level heuristics start with an empty solution and gradually build it, whereas perturbation low-level heuristics start with fully built solutions and change them to generate new solutions. Examples of low-level heuristics are search operators for evolutionary algorithms, local search algorithms and metaheuristics.

\section{Our Proposal: Sentinel}
\label{sec:Sentinel}

Sentinel\footnote{Name inspired by the \textit{Sentinel} character of the X--Men comics by Marvel Comics (\url{http://marvel.com/universe/X-Men}). In the Marvel universe, a Sentinel is a giant robot in charge of killing mutants.} is a multi-objective approach based on hyper-heuristics to automatically generate mutant reduction strategies for Java programs. Instead of trying to directly perform the mutant reduction, Sentinel tries to automatically find a set of optimal strategies that can reduce the cost of mutation testing while maintaining the global mutation score (i.e.\ the mutation score when considering all available mutants).

The ultimate goal of Sentinel is to generate optimal strategies that are effective in reducing mutants. An example of a strategy generated by Sentinel is: A)~retain 80\% of the available operators; B)~execute 100\% of retained operators; C)~group generated mutants by operator; D)~order the groups by number of mutants in descending order; E)~discard the first two groups of mutants; and F)~store in $M'$ 10\% of the remaining mutants at random from each group. At the end of the strategy execution, it will simply output the reduced set of mutants $M'$ to be used as a replacement to the whole set of mutants $M$. These operations -- such as the ones from the example (A--F) -- were extracted from conventional strategies and stored in a grammar used by a GE algorithm~\cite{Ryan1998} (a type of EA) to generate the strategies (explained in the next section). With the GE algorithm and the grammar file, Sentinel is able to combine operations from different kinds of conventional strategies in order to build new and unseen strategies.

The search for strategies is done during a learning phase (i.e., evolutionary process), where Sentinel's GE implementation iteratively generates new strategies by means of recombination of genetic information from parents, mutation of children genes and natural selection of best fit strategies across several subsequent generations (see Section~\ref{sec:Sentinel:Multi-Objective Optimisation}). At the end of the evolutionary process, Sentinel returns a set containing the strategies with the best trade-off between execution time minimisation and mutation score conservation.

For the generation and execution of mutants, Sentinel uses PIT 1.2.0\footnote{\url{http://pitest.org/}}~\cite{Coles2016}, a mutation tool for Java programs. We chose to use PIT because it is one of the fastest and widely used Java mutation tools publicly available~\cite{Bowes2016, Gopinath2016, Laurent2017}. PIT already uses several cost reduction techniques, such as bytecode manipulation, multi-threading, essential operators, test case prioritisation, and code coverage analysis. Furthermore, PIT uses Partial Mutation, meaning that mutants are executed until they are killed, and returning to the user the test cases that were used to kill such mutants (the reduced set of test cases $T'$ mentioned in Section~\ref{sec:Mutant Reduction}). By using Sentinel in combination with PIT, we believe that we can further improve the cost reduction without disregarding the mutation score for non-trivial programs.

In this sense, instead of executing all mutants $M$, the tester chooses a strategy generated by Sentinel and execute it to obtain a reduced set of mutants $M'$ which is used by PIT to perform the mutation analysis. Ideally, the Sentinel training process is done only once and the resulting strategies can be reused every time the tester needs to perform the mutation testing. Therefore, the advantages of Sentinel are: i)~it provides the tester with mutant reduction strategies specially tailored and optimised for their software; and ii)~it automates the generation of strategies, so that a tester does not need to select and configure existing strategies manually through experimentation. The next subsections present details about Sentinel and the generated strategies.

\subsection{Solution Representation and Genotype--Phenotype Mapping}
\label{sec:Representation}

\begin{figure}[tb]
	\centering
	\includegraphics[width=0.8\linewidth]{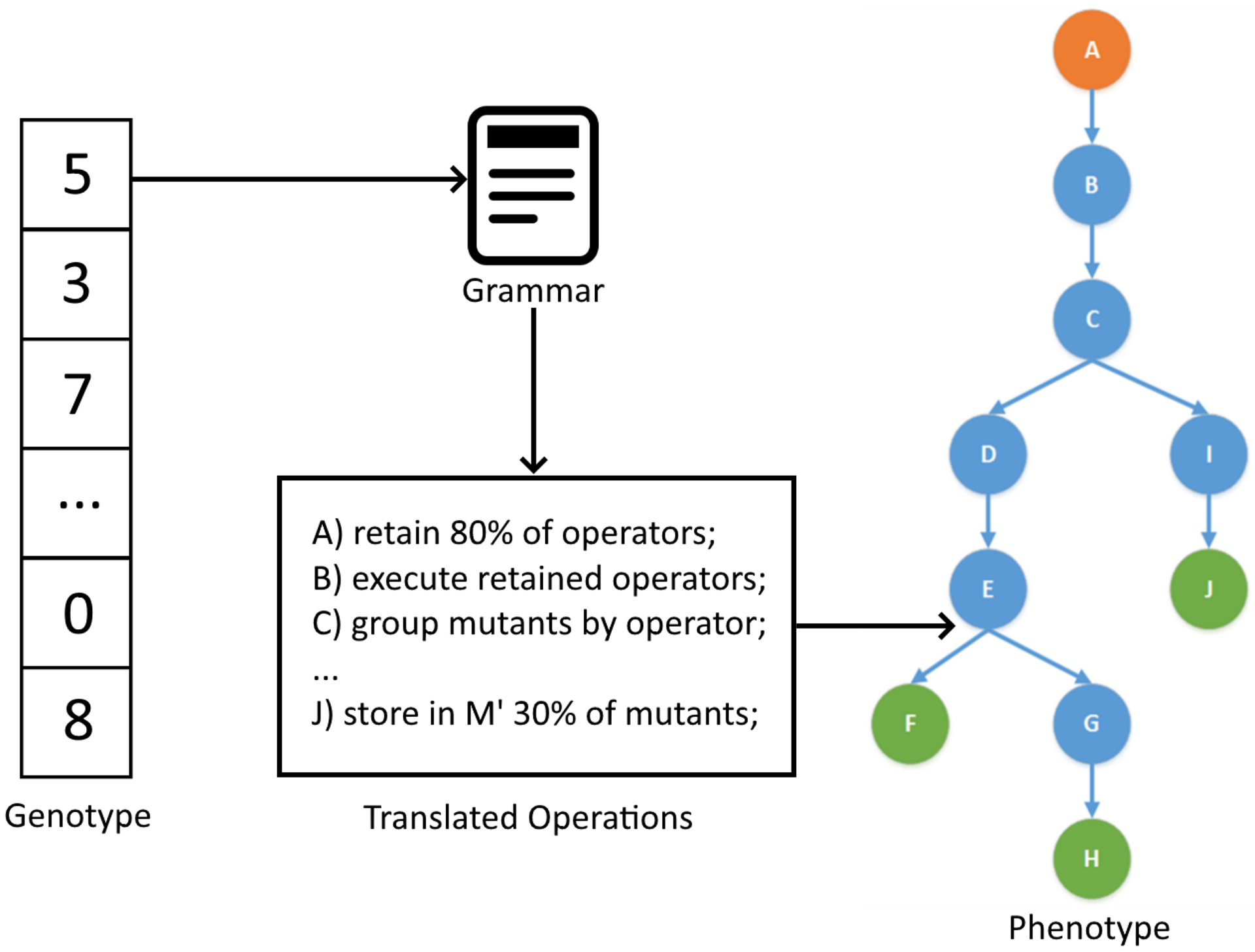}
	\caption{Genotype--Phenotype Mapping of Sentinel.}
	\label{fig:Strategy GPM}
\end{figure}

The overall process for generating a strategy is shown in Figure~\ref{fig:Strategy GPM} and it is called Genotype--Phenotype Mapping (GPM)~\cite{Whigham2017}. 

\begin{figure}[tb]
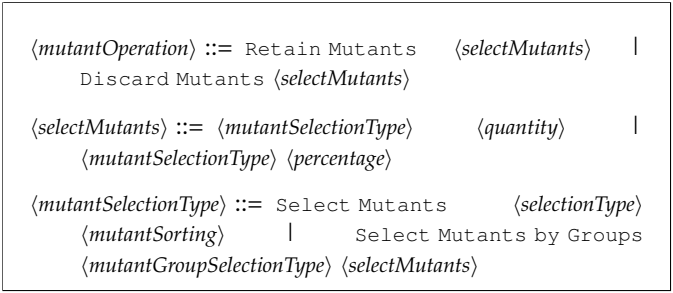

	\centering
	\begin{mdframed}
		\begin{grammar}
			\fontsize{8pt}{8pt}
			<mutantOperation> ::= "Retain Mutants" <selectMutants> | "Discard Mutants" <selectMutants>
			
			<selectMutants> ::= <mutantSelectionType> <quantity> | <mutantSelectionType> <percentage>
			
			<mutantSelectionType> ::= "Select Mutants" <selectionType> <mutantSorting> | "Select Mutants by Groups" <mutantGroupSelectionType> <selectMutants>
		\end{grammar}
	\end{mdframed}
	\caption{Excerpt of Sentinel's grammar.}
	\label{fig:Grammar}
\end{figure}

Although the ultimate goal of Sentinel is evolving a set of executable strategies (i.e., phenotype), GE algorithms~\cite{Ryan1998} cannot work directly on the phenotype and use a chromosome (i.e., genotype) instead to evolve the solutions, similarly to other types of EAs~\cite{Coello2007}. In our case, the chromosome is encoded as an array of $n$ integers, where each integer represents a rule to be selected from a grammar. In fact, the GPM processes transforms this array into a fully executable strategy by using a set of predefined rules forming the grammar. For example, Figure~\ref{fig:Grammar} is an excerpt of the default Sentinel grammar\footnote{The complete grammar used in this study can be found on Sentinel's web page~\cite{SentinelWebsite}.}.

Each one of the rules before $::=$ can be translated into one of the options delimited by $|$. One of the options is chosen by consuming an integer value (gene) of the chromosome: $\mathit{chosen\_option} = \mathit{mod}(\mathit{gene}, \mathit{num\_options})$. If the option has a non-terminal rule between $<$ and $>$, then more genes must be consumed to translate each of those rules into terminal ones. The grammar rules actually represent the available operations which Sentinel encompasses: i)~Retain -- retain mutants or operators, like a filter where only the selected elements remain in the pool; ii)~Discard -- discard mutants or operators; iii)~Group -- group mutants or operators in clusters to perform actions over these groups; or iv)~Execute -- execute the operators to generate mutants. At the end of the GPM process, the genotype genes of a solution are translated into a set of operations for building a strategy (i.e., phenotype). This fully assembled strategy can be seen as an execution tree, where each node is an operation with its required parameters. In the example shown in Figure~\ref{fig:Strategy GPM}, the strategy execution starts at node A, continues until it reaches a leaf node and then proceeds to execute the next branches of the tree until all nodes are executed. During this execution path, the strategy can select operators, execute them to generate mutants and then select the mutants. The final result of such execution is the reduced mutant set $M'$.

\subsection{Objective Functions}
\label{sec:Sentinel:Objective Functions}

Sentinel generates and evaluates strategies according to two objectives: \textit{TIME} and \textit{SCORE}. Both of them are based on the mean values obtained by a strategy $e$ executed for $n$ repetitions over the training instance.

The \textit{TIME} minimisation objective for a given strategy $e$ is given by Equation~\ref{eq:Time}:\begin{equation}
\label{eq:Time}
\downarrow \mathit{\textit{TIME}}(e) = \frac{\sum_{i = 1}^{n} \mathit{cpuTime}(e_{i})}{\sum_{i = 1}^{n} \mathit{cpuTime}(c_i)}
\end{equation}

\noindent where $c$ is the conventional mutation testing procedure (generating all mutants and executing all of them). This function computes the relative CPU time used for executing the strategy $e$ and the resulting mutants in comparison to the conventional mutation procedure, i.e.\ it computes the fraction of time that the strategy takes to execute in comparison to executing all operators and mutants.

Previous work on mutation testing cost reduction usually measure the cost by computing the number of mutants~\cite{Pizzoleto2019}: the fewer the mutants, the cheaper the mutation testing. However, depending on the mutation, a mutant may take more time to execute than another one~\cite{Mresa1999,Zhang2014,Derezinska2016,Derezinska2017,Gligoric2013} (i.e.\ one cannot assume all mutants have the same cost). Thus, in this work we have actually executed each of the mutants selected by a given strategy and computed the time taken to perform the mutation testing activity. It is important to note that, due to the stochastic nature of strategies (both generated by Sentinel and conventional ones from the literature) and to small fluctuations in the CPU time measurement, we adopted a repetition of $n$ strategy executions to compute an average for both functions in each fitness evaluation. The usage of CPU time instead of number of mutants, and the repetition approach are adopted to improve the accuracy of the cost and score measurements of each strategy. Furthermore, we also applied this repetition approach during the experimentation to increase the accuracy in each independent run, while also performing 30 independent runs to cater for the stochasticity and allow the use of statistical tests. The downside of such a measurement is that the experiments become very expensive, both during the training and the testing.

The \textit{SCORE} objective for a strategy $e$ is given by Equation~\ref{eq:Score}:
\begin{equation}
\label{eq:Score}
\uparrow \mathit{\textit{SCORE}}(e) = \frac{\dfrac{1}{n} \sum_{i = 1}^{n} \MS(T'_{i},M)}{\MS(T, M)}
\end{equation}

\noindent where $\MS$ is the mutation score and $T'$ is a test set (selected by PIT from all available test cases $T$) used to kill $M'$. This function computes the average relative test effectiveness obtained by the reduced mutant set, i.e.\ it computes the relative mutation score obtained by a subset of tests $T'$ on all mutants $M$. By maximising this objective, a strategy cannot obtain a mutation score higher than the original one. For instance, if the original mutation score $\MS(T, M)$ is $0.8$ and the average mutation score $\MS(T',M)$ obtained by the strategy $e$ is $0.5$, then \textit{SCORE}$(e)$ will yield $0.625$, since $0.5$ represents 62.5\% of the global mutation score $0.8$. If \textit{SCORE} is $1.0$ ($T' = T_{\mathit{maintain}}$), then it means that $M'$ requires a subset $T'$ that can kill all killable mutants in $M$, i.e\ $T'$ is as good as $T$ in killing $M$ but only $M'$ is required to obtain $T'$.

An optimal strategy is the one that finds a mutant set that is fast to execute and that maintains the global mutation score. However, these two objectives are conflicting because by optimising one, the other will probably worsen (as explained in Section~\ref{sec:Sentinel:Multi-Objective Optimisation}). In the problem of mutant reduction, a strategy that can greatly reduce the cost of the mutation activity at the cost of mutation score approaching can be seen as good as a strategy that greatly approaches the mutation score for a slower execution speed (i.e.\ both are non-dominated). It is up to the engineer to decide which strategy shall be used: the one that saves the most execution time, the one that better approaches the mutation score, or one in between with an acceptable compromise.

\subsection{Implementation}
\label{sec:Sentinel:Implementation Aspects}

Sentinel uses a GE algorithm implementation based on the Non-dominated Sorting Genetic Algorithm II (NSGA--II)~\cite{Deb2002} MOEA to guide the strategy generation. The GE implementation of Sentinel is presented in Algorithm~\ref{alg:GE}. The main structure of the algorithm is very similar to a conventional multi-objective GE, the only exception is that our algorithm uses pruning and duplication operators after the mutation of chromosomes. The former helps to avoid the mutation and crossover of useless genes, and the latter helps to minimise chromosome wraps~\cite{Ryan1998}. The GE parameters used in the experiment are shown in Table~\ref{tab:Parameters} and were chosen based on previous work~\cite{Guizzo2015,Guizzo2017,Mariani2016}.

\begin{algorithm}[tb]
	\caption{GE implementation of Sentinel}
	\label{alg:GE}
	\fontsize{8pt}{8pt}\selectfont
	\Begin{
		$\mathit{chromosomes} \gets$ Randomly initialise the population\;
		$\mathit{strategies} \gets$ Perform the GPM on $\mathit{chromosomes}$ using the grammar file\; \label{alg:GE:mapping}
		$\mathit{time} \gets$ Compute CPU time for the next two steps\;
		$\mathit{mutants} \gets$ Execute $\mathit{strategies}$\; \label{alg:GE:results}
		$\mathit{score} \gets$ Execute $\mathit{mutants}$\;
		Evaluate $\mathit{chromosomes}$ according to $\mathit{time}$ and $\mathit{score}$\; \label{alg:GE:evaluate}
		\While{maximum fitness evaluations not reached}{
			$\mathit{parents} \gets $ Select best parents in $\mathit{chromosomes}$\;
			$\mathit{offspring} \gets$ Recombine $\mathit{parents}$ with crossover\;
			Mutate $\mathit{offspring}$\;
			$\mathit{strategies} \gets$ Perform the GPM on $\mathit{offspring}$ using the grammar file\; \label{alg:GE:mapping offspring}
			$\mathit{time} \gets$ Compute CPU time for the next two steps\;
			$\mathit{mutants} \gets$ Execute $\mathit{strategies}$\;
			$\mathit{score} \gets$ Execute $\mathit{mutants}$\;
			Evaluate $\mathit{chromosomes}$ according to $\mathit{time}$ and $\mathit{score}$\;
			$\mathit{chromosomes} \gets$ Select the best individuals in $\mathit{chromosomes}$ and $\mathit{offspring}$\;
		}
		\Return{Non-dominated strategies in $\mathit{chromosomes}$}\;
	}
\end{algorithm}

\begin{table}[tb]
	\centering
	\caption{Sentinel GE parameters.}
	\label{tab:Parameters}
	\fontsize{6pt}{6pt}\selectfont
	\def\arraystretch{1.2}
	\begin{tabular}{lr}
		\toprule
		Parameter                 &                   Value \\ \midrule
		Independent Runs          &                      30 \\
		Strategy Repetitions      &                       5 \\
		Maximum Evaluations       &                  10,000 \\
		Population Size           &                     100 \\
		Crossover Operator        &  Single Point Crossover \\
		Crossover Probability     &                   100\% \\
		Mutation Operator         & Random Integer Mutation \\
		Mutation Probability      &                     1\% \\
		Prune Probability         &                    10\% \\
		Duplicate Probability     &                    10\% \\
		Lower Gene Bound          &                       0 \\
		Upper Gene Bound          &                     179 \\
		Maximum Chromosome Length &                     100 \\
		Minimum Chromosome Length &                      15 \\
		Maximum Wraps             &                      10 \\ \bottomrule
	\end{tabular}
\end{table}

At the start of the algorithm execution, a set of 100 chromosomes are generated, mapped into strategies and then evaluated. For each strategy, the evaluation consists in: i)~mapping the strategy using GPM; ii)~executing the strategy over a training instance and its test suite; iii)~collecting the reduced mutant set; iv)~executing the mutant set; and v)~computing both \textit{SCORE} and \textit{TIME} for the execution of the strategy and resulting mutants. Then, the best strategies are selected to reproduce and new children are generated throughout crossover and mutation. For the replacement, i.e.\ substitution of parents for their children to survive to the next generation, the NSGA--II procedure of pruning the populations is applied considering both convergence and diversity. This is done for 100 generations and then the algorithm is terminated returning the non-dominated strategies (Pareto front) found along the generations.

In order to reduce the training cost of Sentinel and avoid the execution of all mutants during each evolutionary process, we implemented a caching process to store information about each mutant and operator. In the first step, Sentinel executes each operator and mutant separately for the system. This means that for each mutant Sentinel creates a PIT mutation pool and executes it separately in a different Java Virtual Machine (JVM) instance. Then, the execution cost and mutation score are stored in a cache file, which in turn is used as a source of information during the fitness evaluation. Instead of executing all mutants obtained by a strategy at each fitness evaluation, Sentinel uses the information stored in this cache file to compute the objective values. Even though there is an overhead for setting up a JVM instance for each mutant during the caching, this is more viable than letting the strategies execute all selected mutants. Finally, we use the default options for PIT and apply Sentinel using the seven default mutation operators as commonly done in previous work~\cite{Bowes2016, Coles2016}.

\subsection{Final Remarks on Sentinel}

Sentinel is a novel approach since, as far as we are aware, there is no previous work on automated optimal generation/selection of mutant reduction strategies. Furthermore, Sentinel uses the actual strategies execution time and the score approaching as objectives for the optimisation. It also allows the usage of other objectives, giving the testers the possibility of generating strategies tailored to their needs. Previous work on mutant reduction usually rely only on the number of mutants and mutation score~\cite{Pizzoleto2019}.

\section{Empirical Study Design}
\label{sec:Empirical Study}

This section presents the design of the empirical study conducted to evaluate the strategies generated by Sentinel. The main goal of this evaluation is to determine the capability of the strategies generated by Sentinel in reducing the overall execution time (i.e.\ cost) of mutation testing and in maintaining the mutation score. Therefore, a strategy is better than another if it is better or equal for these two objectives, and better for at least one of them, according to the Pareto dominance concept~\cite{Coello2007}.

\subsection{Research Questions}
\label{RQs}

In order to assess Sentinel's effectiveness we answer three research questions (RQs). 

First of all, we carry out a sanity check to assure that the hyper-heuristic mechanism used by Sentinel effectively learns how to generate good strategies during the training phase and therefore it is better than using a random strategy:

\textbf{RQ1 -- Sanity Check:} Is Sentinel better than a random strategy generation algorithm?

To answer this question, we run both Sentinel and the random generation algorithm for 30 independent runs using 10 different real-world systems widely used in previous work~\cite{Gopinath2016,Laurent2017,Just2014,Zhang2016}. Sentinel is a multi-objective approach, thus each independent run generates an approximated Pareto front where each solution of the front is a generated strategy. Therefore, we evaluate the resulting fronts using the HV and IGD indicators~\cite{Zitzler2003} (described in Section~\ref{sec:Sentinel:Multi-Objective Optimisation}), which have been widely used in previous SBSE work~\cite{FerrucciHRS13,ZhangHJS15,SarroPH16,Sarro2017}.
Because the true Pareto front for the problem we investigate herein is unknown and infeasible to discover, we create such a front with all the non-dominated solutions found by all algorithms, as done in previous work~\cite{Guizzo2015,Mariani2016,SarroPH16,Guizzo2017,Sarro2017}.
For each experiment, we normalise the objective values for computing both indicators, thus the HV and IGD value's range may vary from one experiment to another.

We also used the Kruskal--Wallis statistical test~\cite{Kruskal1952} to assess if there is any statistical difference between the HV and IGD values achieved by the different algorithms, and the Vargha--Delaney A12 effect size~\cite{Vargha2000} to compute the magnitude of the differences. Kruskal--Wallis is a statistical test that can show if the data of two or more groups are statistically different. In this paper we assume the confidence of 95\%, hence if the computed \textit{p-value} is lower than $0.05$, then there is statistical difference between the groups. Kruskal--Wallis is actually an extension of Wilcoxon rank sum test~\cite{Wilcoxon1945} by allowing comparisons between three or more groups. The Vargha--Delaney $\hat{A}_{\mathit{12}}$ effect size gives the magnitude of the difference between two groups A and B. The Vargha--Delaney $\hat{A}_{\mathit{12}}$ value varies between $[0,1]$, where $0.5$ represents absolute no difference between the two groups, values below $0.5$ represent that group B obtains greater values than group A, and values above $0.5$ represent that A obtains greater values than B. We chose these tests because they are non-parametric, i.e.\ they must be used when one cannot assume normal distribution of data. 

After assessing if the strategies generated by Sentinel are better than random guessing in RQ1, we compare these strategies to conventional strategies commonly used in the literature: Selective Mutation (SM), Random Operator Selection (ROS), and Random Mutant Sampling (RMS) (described in Section~\ref{sec:Mutation Testing}). It is important to emphasize that, while RQ1 is designed to evaluate the hyper-heuristic generation process, RQ2 focuses on evaluating the strategies themselves. Therefore, our second question is as follows:

\textbf{RQ2 -- Sentinel's Strategies vs. Conventional Strategies:} Are the strategies generated by Sentinel capable of outperforming conventional mutant reduction strategies in both time reduction and score approaching?

Each type of strategy (SM, ROS, RMS and Sentinel-generated strategies) has multiple strategy variations that differ on their configuration parameters. The conventional strategies were implemented using the same framework as Sentinel and a set of strategies with parameter variations was created for each of them. The strategies are: i)~RMS, randomly selecting from 10\% to 90\% (steps of 10\%) of mutants; ii)~ROS, selecting from 10\% to 90\% (steps of 10\%) of the operators; and iii)~SM, excluding from one to six (steps of one) of the operators that generate the most mutants. RMS is not to be confused with the random generation hyper-heuristic evaluated in RQ1. An RMS strategy selects mutants, whereas a hyper-heuristic generates strategies, thus RQ1 and RQ2 are focused on two different aspects of our experiments.

For each of the 10 systems, we used four different versions. Hence, all strategies were executed over four different versions of the 10 systems with 30 independent runs each, for a total of 1,200 independent runs for each strategy (4,800 in total). We had to limit our analysis to three types of conventional strategies and four system versions due to the high cost of the experiment including manual collection of the data and algorithms running time.

Each of the four types of strategy (Sentinel-generated and the three conventional ones) represents a set of possible solutions, where each strategy is a non-dominated solution in the Pareto front. Given the multiple Pareto fronts, one for each system and type of strategy, we applied the same indicators (HV and IGD) and statistical tests (Kruskal--Wallis and Vargha--Delaney $\hat{A}_{\mathit{12}}$) as the previous RQ to analyse the results. A positive answer to this question will confirm Sentinel's ability to automatically generate better strategies than existing ones.

Once we assessed whether Sentinel is able to provide better strategies than conventional ones, we move to investigating if the strategies learnt by Sentinel on a given version remain effective for subsequent unseen versions, as follows:

\textbf{RQ3 -- Sentinel Strategies Effectiveness Over Time:} Can the strategies generated by Sentinel on a given software version be effectively used to test subsequent versions?

The longer a model remains effective the better. This is a crucial aspect of any automated learner since software change over time and a model built at some point may not be useful later on, thus requiring us to re-train it with consequent costs. If we find that the strategies need to be generated at every new version of the code, as previously generated ones would be ineffective for new versions, then Sentinel would not be practical.

In order to answer RQ3, we investigate this scenario: a tester trains Sentinel by using the current version of a system and, when a new version of the system is available, these previously generated strategies are used (rather than re-run Sentinel on this new version to obtain new strategies). To this end, we use the same data collected for answering RQ2, we use Sentinel to generate strategies for the first version of each of the software systems and then reuse them for the three subsequent versions. We compare the Sentinel-generated strategies to the conventional ones, since the latter can be powerful enough on subsequent versions of the system and the former might not be needed as a consequence.

\subsection{Subjects}
\label{sec:Empirical Study:Subjects}

In our empirical study we used the following 10 real-world systems (also used in previous work~\cite{Gopinath2016,Laurent2017,Just2014,Zhang2016}):

\begin{itemize}
	\item[--] {\it Apache Commons Beanutils}, a library for wrapping around reflection and introspection\footnote{\url{https://github.com/apache/commons-beanutils}};
	\item[--] {\it Apache Commons Codec}, an encoder/decoder for several formats such as Base64 and Hexadecimal\footnote{\url{https://github.com/apache/commons-codec}};
	\item[--] {\it Apache Commons Collections}, a library for collections and arrays manipulation\footnote{\url{https://github.com/apache/commons-collections}};
	\item[--] {\it Apache Commons Lang}, an utility library for \textit{java.lang}\footnote{\url{https://github.com/apache/commons-lang}};
	\item[--] {\it Apache Commons Validator}, a client-server validation library\footnote{\url{https://github.com/apache/commons-validator}};
	\item[--] {\it JFreeChart}, a framework for manipulating charts\footnote{\url{https://github.com/jfree/jfreechart}};
	\item[--] {\it JGraphT}, a library with classes and algorithms for graphs\footnote{\url{https://github.com/jgrapht/jgrapht}};
	\item[--] {\it Joda--Time}, a date and time library\footnote{\url{https://github.com/JodaOrg/joda-time}};
	\item[--] {\it OGNL}, an Object--Graph Navigation Language (OGNL) library to express Java object's properties\footnote{\url{https://github.com/jkuhnert/ognl}};
	\item[--] {\it Wire}, a project for Protocol Buffers for Android\footnote{\url{https://github.com/square/wire}}.
\end{itemize}

Because these programs have several minor and major versions and it would be impracticable to test Sentinel on all of their versions, we decided to use only the oldest major versions of each program that could be mutated by PIT for the training phase and three subsequent versions for the testing phase. The properties of these programs are summarised in Table~\ref{tab:Programs}. The code churn is computed by summing the number of new Logical Lines Of Code (LLOC), modified LLOC, and removed LLOC.

\begin{table}[tb]
	\caption{\textbf{Software versions used in our study.} \textit{LLOC} is the number of logical lines of code; \textit{Churn} is the code churn from previous version; $|T|$ is the number of test cases in the test suite; \textit{T. LLOC} is the number of logical lines of test code; \textit{Cov} is the statement and branch coverage percentages; $|M|$ is the number of mutants generated by PIT; \textit{MS} is the mutation score obtained with the available test suite.}
	\label{tab:Programs}
	\fontsize{6pt}{6pt}\selectfont
	\def\arraystretch{1.2}
	\begin{tabular}{lrrrrrrr}
		\toprule
		Program           &   \textit{LLOC} &  \textit{Churn} & $|T|$ & \textit{T. LLOC} & \textit{Cov} & $|M|$ & $\MS$ \\ \midrule
		beanutils-1.8.0   & 11,279 &     -- &   877 &  20,486 & 60/62 &  2,827 & 0.61 \\
		beanutils-1.8.1   & 11,362 &    179 &   892 &  20,934 & 60/61 &  2,855 & 0.59 \\
		beanutils-1.8.2   & 11,362 &     22 &   893 &  20,966 & 60/61 &  2,856 & 0.59 \\
		beanutils-1.8.3   & 11,376 &     35 &   896 &  21,033 & 60/61 &  2,857 & 0.59 \\ \hline
		codec-1.4         &  2,994 &     -- &   284 &   6,846 & 99/94 &  1,587 & 0.91 \\
		codec-1.5         &  3,551 &  2,631 &   380 &   8,307 & 98/93 &  1,895 & 0.91 \\
		codec-1.6         &  4,554 &  1,472 &   421 &   9,034 & 96/93 &  2,196 & 0.88 \\
		codec-1.11        &  8,109 &  7,858 &   875 &  11,774 & 96/90 &  3,473 & 0.85 \\ \hline
		collections-3.0   & 22,842 &     -- & 1,896 &  22,828 & 90/74 &  7,488 & 0.29 \\
		collections-3.1   & 25,372 &  6,552 & 2,346 &  25,734 & 82/75 &  8,298 & 0.30 \\
		collections-3.2   & 26,323 &  2,022 & 2,566 &  29,076 & 81/75 &  8,637 & 0.31 \\
		collections-3.2.1 & 26,323 &    161 & 2,566 &  29,076 & 81/75 &  8,632 & 0.31 \\ \hline
		lang-3.0          & 18,997 &     -- & 1,902 &  31,008 & 94/91 &  9,072 & 0.85 \\
		lang-3.0.1        & 19,495 &    758 & 1,964 &  31,804 & 92/90 &  9,328 & 0.85 \\
		lang-3.1          & 19,499 &    354 & 1,976 &  32,446 & 92/90 &  9,333 & 0.85 \\
		lang-3.2          & 22,532 & 12,053 & 2,390 &  38,963 & 94/90 & 10,970 & 0.86 \\ \hline
		validator-1.4.0   &  5,411 &     -- &   414 &   6,367 & 79/72 &  1,811 & 0.74 \\
		validator-1.4.1   &  6,031 &  1,228 &   442 &   7,389 & 82/74 &  1,917 & 0.75 \\
		validator-1.5.0   &  6,669 &  1,390 &   481 &   7,922 & 84/74 &  1,979 & 0.75 \\
		validator-1.5.1   &  7,014 &    524 &   486 &   8,051 & 85/75 &  1,982 & 0.75 \\ \hline
		jfreechart-1.0.0  & 68,796 &     -- & 1,023 &  26,823 & 62/39 & 23,417 & 0.27 \\
		jfreechart-1.0.1  & 68,663 &  1,816 & 1,027 &  27,016 & 62/39 & 23,490 & 0.28 \\
		jfreechart-1.0.2  & 73,162 &  9,036 & 1,073 &  28,504 & 63/39 & 24,091 & 0.28 \\
		jfreechart-1.0.3  & 77,621 & 12,640 & 1,234 &  32,825 & 47/40 & 26,401 & 0.28 \\ \hline
		jgrapht-0.9.0     & 12,978 &     -- &   188 &   7,030 & 79/72 &  2,976 & 0.62 \\
		jgrapht-0.9.1     & 13,822 &  2,110 &   647 &   8,184 & 79/73 &  3,147 & 0.66 \\
		jgrapht-0.9.2     & 15,661 &  8,079 &   728 &  10,046 & 80/75 &  3,825 & 0.69 \\
		jgrapht-1.0.0     & 16,417 & 11,442 & 1,201 &  13,609 & 81/75 &  4,270 & 0.71 \\ \hline
		joda-2.8          & 28,479 &     -- & 2,967 &  54,645 & 89/81 & 10,225 & 0.62 \\
		joda-2.8.1        & 28,479 &    107 & 2,967 &  54,645 & 89/81 & 10,225 & 0.62 \\
		joda-2.8.2        & 28,479 &    142 & 2,967 &  54,645 & 89/81 & 10,225 & 0.62 \\
		joda-2.9          & 28,624 &    366 & 2,985 &  54,985 & 89/81 & 10,321 & 0.62 \\ \hline
		ognl-3.1          & 16,103 &     -- &    54 &   6,229 & 69/60 &  5,650 & 0.25 \\
		ognl-3.1.1        & 16,102 &     27 &    53 &   6,223 & 69/60 &  5,652 & 0.26 \\
		ognl-3.1.2        & 16,103 &      9 &    56 &   6,252 & 69/60 &  5,652 & 0.26 \\
		ognl-3.1.3        & 16,109 &      9 &    57 &   6,268 & 69/60 &  5,654 & 0.26 \\ \hline
		wire-2.0.0        &  1,354 &     -- &    70 &   1,776 & 59/53 &    513 & 0.64 \\
		wire-2.0.1        &  1,354 &      0 &    70 &   1,776 & 59/53 &    513 & 0.64 \\
		wire-2.0.2        &  1,353 &      3 &    70 &   1,776 & 59/53 &    513 & 0.64 \\
		wire-2.0.3        &  1,405 &    226 &    71 &   1,794 & 58/53 &    524 & 0.62 \\ \bottomrule
	\end{tabular}
\end{table}

\subsection{Threats to Validity}
\label{sec:Empirical Study:Threats to Validity}

We evaluated the effectiveness of the strategies generated by Sentinel using 10 different software programs having four different versions each. The initial experiments design was to use 10 systems and 10 versions each for a total of 100 projects. However, using such a large number of projects revealed to be infeasible, given the high cost of computing the CPU time of each strategy execution. Even though the total number system versions evaluated (40) is similar or even higher to the number of systems in related work~\cite{Gopinath2016,Just2014, Laurent2017,Zhang2016}, we cannot assert that this is enough to generalise the results to other systems. Another threat to the generalisation of the results is the churn between versions which, for some systems, are somewhat small. To minimise these threats, we tried to evaluate systems of different sizes, domains, test coverage, mutation score, and that had already been used in previous work.

To answer RQ2 we trained Sentinel on the first version of a program and tested the generated strategies on the same version and three unseen subsequent versions. In this way we recreated a scenario that we would expect a tester to follow when using Sentinel: train on a version of their software, and use the generated strategies while testing that same version and subsequent ones. Training and testing an approach on the same version may usually lead to overfitting. In order to minimise this threat, we consider different versions of the software and we perform an analysis using only these subsequent versions in RQ3.

Because Sentinel uses PIT, the results of the experiments depend on the configuration of PIT and its internal features, such as test case prioritisation and mutation operator execution. In this sense, using different configurations for PIT may lead to different results, specially considering that the set $T'$ used to compute the mutation score is obtained based on PIT's test case prioritisation. To minimise this threat, we left all the default configurations of PIT unchanged to all strategy types, so that they all are tested in a same environment. Furthermore, we performed 30 independent runs to cater for stochasticity.

We compared Sentinel-generated strategies to all identified variations of the SM and ROS strategies for the default operators of PIT. For RMS we had to limit the number of strategies by defining a fixed step percentage for their configuration as done in previous work~\cite{Zhang2013,Lima2016,Zhang2010,Mathur1994}. Even though the step sizes vary from one work to another, limiting the set of sampling strategies is done to allow the empirical experimentation in feasible time. Using a smaller step of 1\% instead of 10\% for RMS could yield different results. Furthermore, the strategies generated by Sentinel are tailored specifically for the system in hand. These strategies are potentially more powerful than conventional ones for the system, hence the good results may be biased towards Sentinel's strategies.

\section{Empirical Study Results}
\label{sec:Empirical Study:Results}

This section presents and discusses the results obtained in our experiment. Subsection~\ref{sec:Empirical Study:Results:RQ1} presents the results regarding RQ1, which compares Sentinel vs. a Random hyper-heuristic generation. Subsection~\ref{sec:Empirical Study:Results:RQ2} presents the results regarding RQ2, which compares the strategies generated by Sentinel with three common strategies from the literature. Subsection~\ref{sec:Empirical Study:Results:RQ3} presents the results answering RQ3, which assesses the effectiveness of Sentinel's strategies over time. Finally, in Section~\ref{sec:Empirical Study:Results:Discussion} we discuss some final observations about the study.

\subsection{RQ1 -- Sentinel vs. Random Hyper-Heuristic}
\label{sec:Empirical Study:Results:RQ1}

Table~\ref{tab:Indicator Results RQ1} shows the mean HV and IGD results (standard deviation in brackets) obtained comparing Sentinel vs. Random hyper-heuristic, together with the corresponding effect sizes. The best indicator values and \textit{p-values} lower than $0.05$ are highlighted in bold. For the HV indicator, an effect size value closer to $1$ is favourable to Sentinel (i.e.\ greater HV values more often), whereas an effect size value closer to $0$ for IGD is better for Sentinel (i.e.\ lower IGD values more often).

\begin{table}[tb]
	\centering
	\caption{\textbf{RQ1}: HV and IGD mean results of Sentinel vs. Random Heuristic (standard deviation in brackets). Greater HV values and lower IGD values are better (best results in bold).
	}
	\label{tab:Indicator Results RQ1}
	\fontsize{6pt}{6pt}\selectfont
	\def\arraystretch{1.2}
	\begin{tabular}{llrrrr}
		\toprule
		Program          & Ind. &                   Sentinel &            Random &  \textit{p-value} &        Effect size \\ \midrule
		beanutils-1.8.0  & HV   &      \textbf{0.87 (0.001)} &      0.83 (0.004) & \textbf{2.87E-11} &     \textbf{1 (L)} \\
		                 & IGD  & \textbf{1.98E-4 (6.59E-6)} & 3.24E-4 (3.58E-5) & \textbf{2.87E-11} &     \textbf{0 (L)} \\ \hline
		codec-1.4        & HV   &      \textbf{0.91 (0.001)} &      0.89 (0.002) & \textbf{2.87E-11} &     \textbf{1 (L)} \\
		                 & IGD  & \textbf{2.58E-4 (8.78E-5)} & 3.13E-4 (4.89E-5) &    \textbf{0.003} &           0.29 (M) \\ \hline
		collections-3.0  & HV   &      \textbf{0.83 (0.004)} &      0.81 (0.003) & \textbf{2.87E-11} &     \textbf{1 (L)} \\
		                 & IGD  & \textbf{2.32E-4 (4.05E-5)} & 3.31E-4 (3.78E-5) &  \textbf{1.48E-9} &  \textbf{0.05 (L)} \\ \hline
		lang-3.0         & HV   &      \textbf{0.88 (0.002)} &      0.86 (0.003) & \textbf{2.87E-11} &     \textbf{1 (L)} \\
		                 & IGD  & \textbf{1.83E-4 (1.49E-5)} & 3.08E-4 (3.58E-5) & \textbf{2.87E-11} &     \textbf{0 (L)} \\ \hline
		validator-1.4.0  & HV   &      \textbf{0.90 (0.002)} &      0.87 (0.005) & \textbf{2.87E-11} &     \textbf{1 (L)} \\
		                 & IGD  & \textbf{2.11E-4 (1.37E-5)} & 4.12E-4 (3.60E-5) & \textbf{2.87E-11} &     \textbf{0 (L)} \\ \hline
		jfreechart-1.0.0 & HV   &      \textbf{0.92 (0.002)} &       0.89 (0.01) & \textbf{2.87E-11} &     \textbf{1 (L)} \\
		                 & IGD  & \textbf{2.36E-4 (4.04E-5)} & 5.48E-4 (9.08E-5) & \textbf{3.17E-11} & \textbf{0.001 (L)} \\ \hline
		jgrapht-0.9.0    & HV   &      \textbf{0.93 (0.001)} &       0.90 (0.05) & \textbf{2.87E-11} &     \textbf{1 (L)} \\
		                 & IGD  & \textbf{2.78E-4 (7.15E-5)} & 4.96E-4 (5.47E-5) & \textbf{1.15E-10} &  \textbf{0.02 (L)} \\ \hline
		joda-time-2.8    & HV   &      \textbf{0.85 (0.002)} &      0.82 (0.004) & \textbf{2.87E-11} &     \textbf{1 (L)} \\
		                 & IGD  & \textbf{1.88E-4 (1.64E-5)} & 3.62E-4 (3.46E-5) & \textbf{2.87E-11} &     \textbf{0 (L)} \\ \hline
		ognl-3.1         & HV   &      \textbf{0.97 (0.002)} &      0.95 (0.003) & \textbf{2.87E-11} &     \textbf{1 (L)} \\
		                 & IGD  & \textbf{5.15E-4 (1.64E-4)} & 6.18E-4 (1.09E-4) &    \textbf{0.004} &           0.29 (M) \\ \hline
		wire-2.0.0       & HV   &      \textbf{0.92 (0.004)} &      0.90 (0.004) & \textbf{2.87E-11} &     \textbf{1 (L)} \\
		                 & IGD  & \textbf{4.89E-4 (1.39E-4)} & 4.89E-4 (5.99E-5) &              0.35 &           0.43 (N) \\ \bottomrule
	\end{tabular}
\end{table}

We can observe that Sentinel is able to obtain better results than the Random hyper-heuristic for all programs and for both quality indicators with the only exception of IGD for {\it wire-2.0.0} where the results are similar. Moreover, the Kruskal--Wallis statistical test confirms statistical difference with 95\% confidence (\textit{p-value} $ < 0.05$) for all comparisons and the Vargha--Delaney $\hat{A}_{\mathit{12}}$ effect size shows large differences in 17 out of 20 comparisons in favour of Sentinel, with only two medium and one negligible differences for the IGD comparisons. As mentioned, the only equivalence in these results is for IGD for the smallest system \textit{wire-2.0.0}, where Kruskal--Wallis yielded no difference with a negligible Vargha--Delaney $\hat{A}_{\mathit{12}}$ effect size.

These results show that Sentinel is capable of generating more effective strategies than a Random hyper-heuristic. Therefore, we can positively answer our first research question:

~

\noindent
\fbox{\begin{minipage}{0.98 \columnwidth}
\textbf{RQ1}: \textit{The strategies generated by Sentinel are better than the strategies generated by a random hyper-heuristic.}
\end{minipage}
}

\subsection{RQ2 -- Sentinel's Strategies vs. Conventional Strategies}
\label{sec:Empirical Study:Results:RQ2}

\begin{figure*}[tb]
	\begin{subfigure}{0.5\linewidth}
		\centering
		\includegraphics[width=0.8\linewidth]{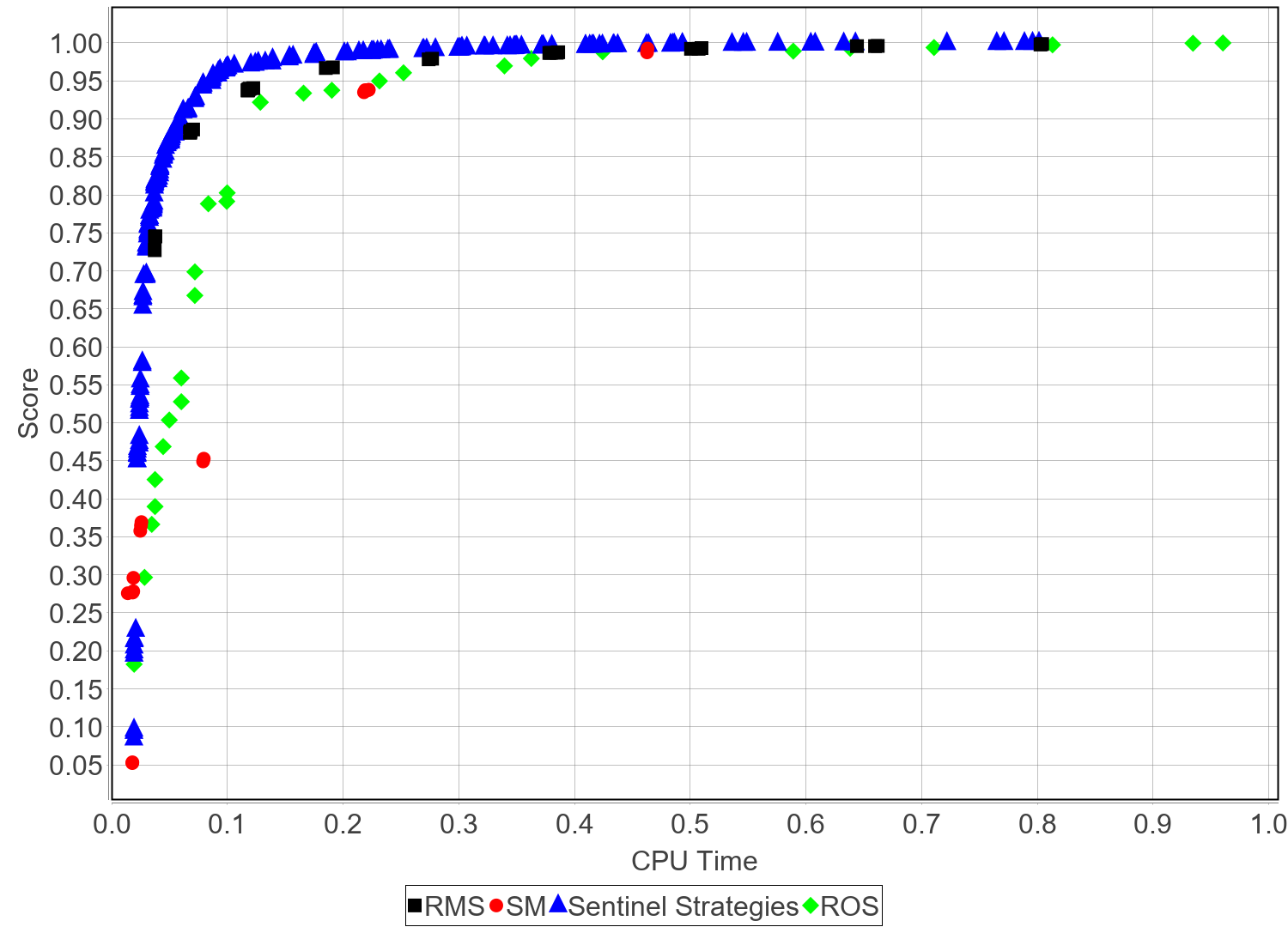}
		\caption{{\it jfreechart-1.0.0}}
	\end{subfigure}
	\begin{subfigure}{0.5\linewidth}
		\centering
		\includegraphics[width=0.8\linewidth]{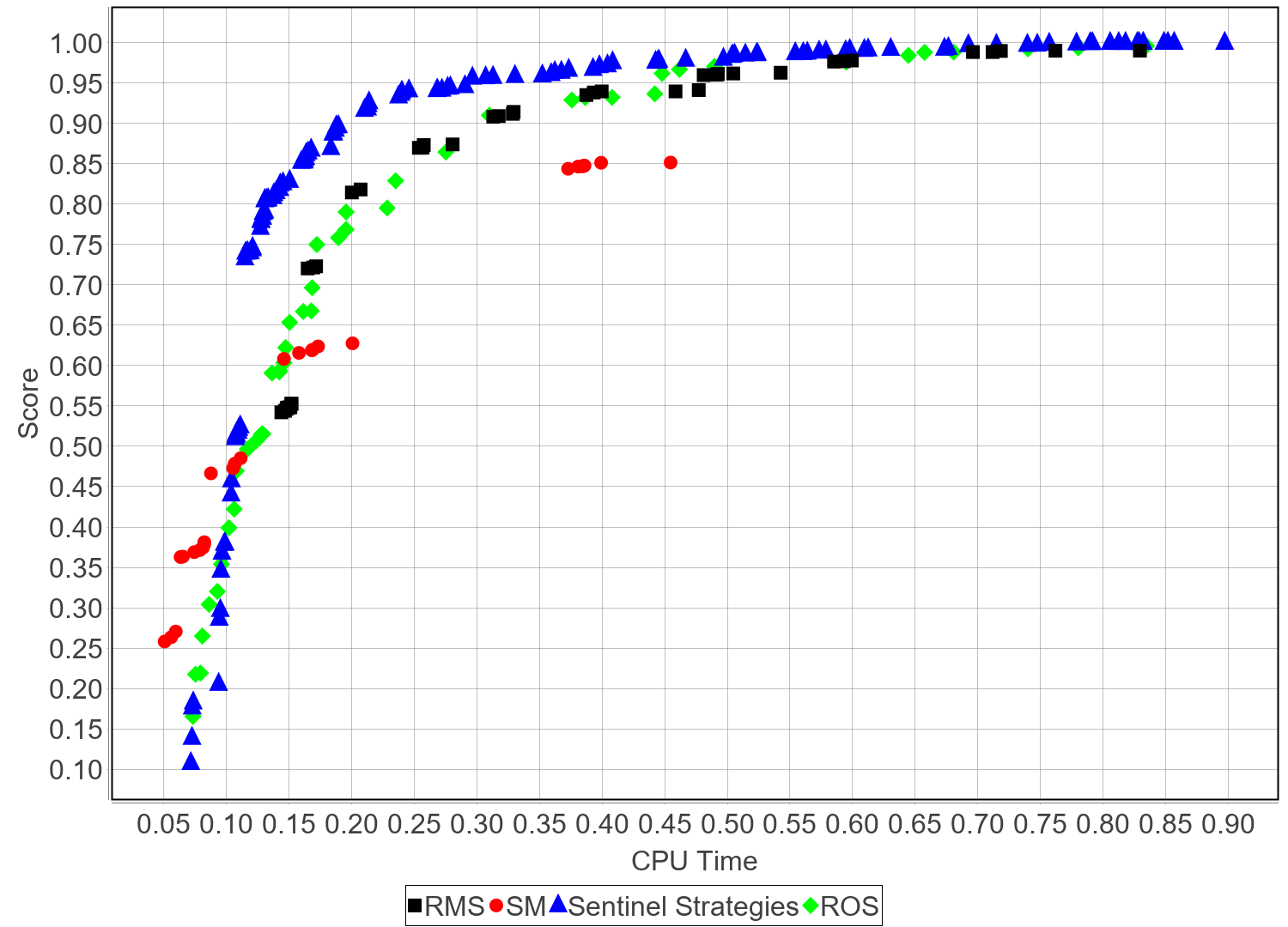}
		\caption{{\it joda-time-2.8}}
	\end{subfigure}
	\caption{Pareto fronts comparing the Mutation Score Approaching (y-axis) vs. CPU time (x-axis) of the strategies obtained by Sentinel and the conventional strategies RMS, SM, and ROS for {\it jfreechart-1.0.0} and {\it joda-time-2.8}. The global mutation score is maintained as the original during the score approaching computation, hence a score of 1.0 in the y-axis means that the strategy obtained the same mutation score as when using the whole set of mutants and test cases.}
	\label{fig:Pareto}
\end{figure*}

As discussed in Section~\ref{sec:Sentinel:Multi-Objective Optimisation}, evaluating each objective (\textit{SCORE} and \textit{TIME}) separately is not meaningful, because in a multi-objective problem, the trade-off between the objectives is what really depicts the results quality. Evaluating each objective separately can be misleading, because a strategy that is good in one objective can be significantly worse in the other\footnote{For completeness, we report the average approaching mutation score  and relative CPU time for Sentinel strategies, and the conventional strategies  in Table~\ref{tab:TimeandScore}. However, care should be taken when interpreting these results separately, because, as explained in Section~\ref{sec:Sentinel:Multi-Objective Optimisation}, both objectives are equally important.}. What an engineer should do in this situation is to look at the non-dominated solutions in the objective space and select the one that best fits their needs. Figure~\ref{fig:Pareto} depicts the Pareto fronts of the non-dominated strategies tested in 30 independent runs for \textit{joda-time-2.8} and \textit{jfreechart-1.0.0} (two of the biggest systems in our experiment) to exemplify this trade-off.

We can observe that Sentinel strategies are more scattered along the Pareto curve and are generally closer to the top left corner of the chart, which indicates a better trade-off between the objectives and generally more non-dominated solutions\footnote{The resulting fronts for the other 38 systems are similar to this, but for space reasons we omitted them from the paper. They are available at Sentinel's website~\cite{SentinelWebsite}.}. Even though SM strategies have a lower average execution time (concentrated on the left half of the plots) and RMS have a greater mutation score approaching (top half of the plots), they are usually dominated by strategies generated by Sentinel. In other words, when using a strategy generated by Sentinel, there is a greater chance of this strategy yielding a lower execution time for the same or better mutation score approaching than conventional strategies, or a better mutation score approaching for the same or better CPU time. For instance, if we consider a minimum mutation score approaching of 0.9 when selecting a strategy for \textit{jfreechart}, the cheapest Sentinel strategy would yield a reduction of approximately 93.5\% in execution time, RMS would yield $\approx$88\%, ROS $\approx$87\%, and SM $\approx$77\%.

We can quantify this trade-off and objectively evaluate the results considering both objectives (CPU time and score approaching) at the same time by computing the HV and IGD indicators (defined in Section~\ref{sec:Sentinel:Multi-Objective Optimisation}). Using such quality indicators for comparing sets of strategies is crucial, because if we report the optimisation of only one objective, then it would not say much about the trade-off with the other. By following the standard procedure of multi-objective analysis of reporting quality indicator values, we can assess the overall performance of the algorithms. In our case, the greater the HV of a resulting set of strategies, the greater the probability the strategies in this set have in outperforming the strategies in other sets with respect to the trade-off of score and time. Moreover, a set of strategies with an IGD value close to $0$ contains strategies that are very close to the best ones found for the program.

We report the HV and IGD results in Tables~\ref{tab:HV Results} and~\ref{tab:IGD Results} respectively. The standard deviation is shown in parenthesis. The best values and \textit{p-values} lower than $0.05$ are highlighted in bold. If two or more indicator values are highlighted in bold in the same row, it means that the statistical test showed no difference between the best strategy and the other highlighted strategies for a given system.

\begin{table}[tb]
	\centering
	\caption{\textbf{RQ2--RQ3:} HV results of Sentinel-generated strategies vs. conventional strategies (standard deviation in brackets). Greater HV values are better (best in bold).}
	\label{tab:HV Results}
	\fontsize{6pt}{6pt}\selectfont
	\def\arraystretch{1.2}
	\begin{tabular}{lrrrrr}
		\toprule
		Program           &              Sentinel &                   SM &          RMS &         ROS &              p-value \\ \midrule
		beanutils-1.8.0   &  \textbf{0.81 (0.02)} &          0.79 (0.02) &  0.76 (0.03) & 0.77 (0.04) &     \textbf{3.3E-10} \\
		beanutils-1.8.1   &  \textbf{0.79 (0.01)} & \textbf{0.79 (0.03)} &  0.73 (0.02) & 0.75 (0.04) &     \textbf{1.6E-11} \\
		beanutils-1.8.2   &  \textbf{0.79 (0.02)} & \textbf{0.78 (0.03)} &  0.74 (0.03) & 0.76 (0.04) &     \textbf{1.9E-09} \\
		beanutils-1.8.3   &  \textbf{0.78 (0.03)} & \textbf{0.78 (0.03)} &  0.73 (0.03) & 0.75 (0.04) &     \textbf{1.9E-10} \\ \hline
		codec-1.4         &  \textbf{0.89 (0.01)} &          0.87 (0.02) & 0.78 (0.005) & 0.84 (0.02) & \textbf{$<$ 2.2E-16} \\
		codec-1.5         & \textbf{0.90 (0.005)} & \textbf{0.89 (0.01)} & 0.80 (0.005) & 0.84 (0.02) & \textbf{$<$ 2.2E-16} \\
		codec-1.6         & \textbf{0.88 (0.003)} & \textbf{0.89 (0.02)} & 0.78 (0.004) & 0.82 (0.02) & \textbf{$<$ 2.2E-16} \\
		codec-1.11        &  \textbf{0.91 (0.01)} & \textbf{0.89 (0.01)} &  0.83 (0.01) & 0.83 (0.03) & \textbf{$<$ 2.2E-16} \\\hline
		collections-3.0   &  \textbf{0.92 (0.01)} &          0.74 (0.01) &  0.86 (0.01) & 0.79 (0.03) & \textbf{$<$ 2.2E-16} \\
		collections-3.1   &  \textbf{0.89 (0.01)} &          0.72 (0.02) &  0.84 (0.01) & 0.77 (0.02) & \textbf{$<$ 2.2E-16} \\
		collections-3.2   &  \textbf{0.90 (0.01)} &          0.73 (0.02) &  0.84 (0.01) & 0.77 (0.02) & \textbf{$<$ 2.2E-16} \\
		collections-3.2.1 &  \textbf{0.90 (0.01)} &          0.73 (0.01) &  0.85 (0.01) & 0.77 (0.02) & \textbf{$<$ 2.2E-16} \\\hline
		lang-3.0          & \textbf{0.95 (0.004)} &          0.88 (0.01) &  0.90 (0.01) & 0.85 (0.02) & \textbf{$<$ 2.2E-16} \\
		lang-3.0.1        & \textbf{0.94 (0.005)} &          0.86 (0.01) &  0.90 (0.01) & 0.84 (0.02) & \textbf{$<$ 2.2E-16} \\
		lang-3.1          &  \textbf{0.94 (0.01)} &          0.86 (0.02) &  0.90 (0.01) & 0.84 (0.02) & \textbf{$<$ 2.2E-16} \\
		lang-3.2          &  \textbf{0.95 (0.01)} &          0.87 (0.02) &  0.91 (0.01) & 0.86 (0.02) & \textbf{$<$ 2.2E-16} \\\hline
		validator-1.4.0   & \textbf{0.86 (0.004)} &          0.79 (0.01) & 0.75 (0.005) & 0.76 (0.02) & \textbf{$<$ 2.2E-16} \\
		validator-1.4.1   &  \textbf{0.87 (0.01)} &          0.80 (0.02) &  0.76 (0.01) & 0.75 (0.02) & \textbf{$<$ 2.2E-16} \\
		validator-1.5.0   & \textbf{0.86 (0.004)} &          0.77 (0.01) &  0.74 (0.01) & 0.74 (0.03) & \textbf{$<$ 2.2E-16} \\
		validator-1.5.1   &  \textbf{0.85 (0.01)} &          0.76 (0.02) &  0.74 (0.01) & 0.74 (0.02) & \textbf{$<$ 2.2E-16} \\\hline
		jfreechart-1.0.0  & \textbf{0.97 (0.001}) &         0.84 (0.001) & 0.94 (0.001) & 0.84 (0.02) & \textbf{$<$ 2.2E-16} \\
		jfreechart-1.0.1  & \textbf{0.97 (0.003)} &          0.85 (0.01) & 0.94 (0.005) & 0.85 (0.02) & \textbf{$<$ 2.2E-16} \\
		jfreechart-1.0.2  & \textbf{0.97 (0.004)} &          0.85 (0.01) &  0.94 (0.01) & 0.85 (0.02) & \textbf{$<$ 2.2E-16} \\
		jfreechart-1.0.3  & \textbf{0.97 (0.003)} &          0.86 (0.01) & 0.94 (0.005) & 0.85 (0.02) & \textbf{$<$ 2.2E-16} \\\hline
		jgrapht-0.9.0     &  \textbf{0.91 (0.01)} &          0.86 (0.01) &  0.86 (0.01) & 0.84 (0.02) & \textbf{$<$ 2.2E-16} \\
		jgrapht-0.9.1     &  \textbf{0.91 (0.01)} &          0.87 (0.01) &  0.85 (0.01) & 0.83 (0.02) & \textbf{$<$ 2.2E-16} \\
		jgrapht-0.9.2     & \textbf{0.90 (0.004)} &          0.87 (0.01) &  0.85 (0.01) & 0.87 (0.02) & \textbf{$<$ 2.2E-16} \\
		jgrapht-1.0.0     & \textbf{0.89 (0.005)} &          0.87 (0.02) &  0.85 (0.01) & 0.87 (0.01) & \textbf{$<$ 2.2E-16} \\\hline
		joda-time-2.8     &  \textbf{0.91 (0.01)} &          0.72 (0.01) &  0.83 (0.01) & 0.80 (0.02) & \textbf{$<$ 2.2E-16} \\
		joda-time-2.8.1   & \textbf{0.90 (0.005)} &          0.71 (0.01) &  0.82 (0.01) & 0.79 (0.02) & \textbf{$<$ 2.2E-16} \\
		joda-time-2.8.2   & \textbf{0.90 (0.005)} &          0.71 (0.01) &  0.81 (0.01) & 0.78 (0.02) & \textbf{$<$ 2.2E-16} \\
		joda-time-2.9     &  \textbf{0.90 (0.01)} &          0.70 (0.01) &  0.81 (0.01) & 0.78 (0.02) & \textbf{$<$ 2.2E-16} \\\hline
		ognl-3.1          & \textbf{0.99 (0.003)} &         0.98 (0.003) &  0.97 (0.01) & 0.92 (0.02) & \textbf{$<$ 2.2E-16} \\
		ognl-3.1.1        & \textbf{0.99 (0.003)} &          0.97 (0.01) &  0.92 (0.01) & 0.92 (0.03) & \textbf{$<$ 2.2E-16} \\
		ognl-3.1.2        & \textbf{0.99 (0.001)} &          0.97 (0.01) & 0.97 (0.002) & 0.92 (0.02) & \textbf{$<$ 2.2E-16} \\
		ognl-3.1.3        & \textbf{0.98 (0.001)} &          0.96 (0.01) & 0.91 (0.002) & 0.91 (0.02) & \textbf{$<$ 2.2E-16} \\\hline
		wire-2.0.0        &           0.80 (0.01) & \textbf{0.83 (0.01)} &  0.62 (0.01) & 0.80 (0.01) & \textbf{$<$ 2.2E-16} \\
		wire-2.0.1        &           0.84 (0.02) & \textbf{0.87 (0.01)} &  0.65 (0.02) & 0.84 (0.02) & \textbf{$<$ 2.2E-16} \\
		wire-2.0.2        &           0.83 (0.01) & \textbf{0.86 (0.01)} &  0.64 (0.01) & 0.83 (0.02) & \textbf{$<$ 2.2E-16} \\
		wire-2.0.3        &           0.66 (0.01) & \textbf{0.74 (0.05)} &  0.51 (0.01) & 0.67 (0.01) & \textbf{$<$ 2.2E-16} \\ \bottomrule
	\end{tabular}
\end{table}
\begin{table*}[tb]
	\centering
	\caption{\textbf{RQ2--RQ3:} IGD results of Sentinel-generated strategies vs. conventional strategies (standard deviation in brackets). Lower IGD values are better (best values in bold).}
	\label{tab:IGD Results}
	\fontsize{6pt}{6pt}\selectfont
	\def\arraystretch{1.2}
	\newcolumntype{R}[1]{>{\raggedleft\arraybackslash}p{#1}}
	\begin{tabular}{lrrrrr}
		\toprule
		Program           &                   Sentinel &                SM &               RMS &               ROS &              p-value \\ \midrule
		beanutils-1.8.0   & \textbf{1.61E-4 (6.32E-6)} & 2.05E-3 (2.45E-4) & 2.17E-3 (8.80E-5) & 2.02E-3 (5.73E-4) &     \textbf{2.1E-15} \\
		beanutils-1.8.1   & \textbf{1.71E-4 (7.66E-6)} & 2.25E-3 (2.86E-4) & 2.20E-3 (6.13E-5) & 2.10E-3 (5.49E-4) &     \textbf{1.2E-14} \\
		beanutils-1.8.2   & \textbf{1.94E-4 (1.51E-5)} & 2.24E-3 (3.25E-4) & 2.20E-3 (1.00E-4) & 2.20E-3 (4.81E-4) &     \textbf{1.8E-14} \\
		beanutils-1.8.3   & \textbf{1.89E-4 (1.45E-5)} & 2.22E-3 (3.05E-4) & 2.18E-3 (7.09E-5) & 2.19E-3 (5.11E-4) &     \textbf{1.2E-14} \\\hline
		codec-1.4         & \textbf{2.03E-4 (9.83E-6)} & 2.57E-3 (1.57E-4) & 2.95E-3 (8.15E-5) & 2.94E-3 (3.48E-4) & \textbf{$<$ 2.2E-16} \\
		codec-1.5         & \textbf{1.70E-4 (6.53E-6)} & 2.78E-3 (7.13E-5) & 3.40E-3 (8.30E-5) & 3.44E-3 (3.17E-4) & \textbf{$<$ 2.2E-16} \\
		codec-1.6         & \textbf{1.65E-4 (3.44E-6)} & 2.67E-3 (1.04E-4) & 3.49E-3 (6.22E-5) & 3.55E-3 (4.64E-4) & \textbf{$<$ 2.2E-16} \\
		codec-1.11        & \textbf{1.57E-4 (6.48E-6)} & 2.19E-3 (4.29E-5) & 3.15E-3 (4.93E-5) & 3.03E-3 (3.93E-4) & \textbf{$<$ 2.2E-16} \\\hline
		collections-3.0   & \textbf{1.73E-4 (1.45E-5)} & 1.73E-3 (2.34E-5) & 2.31E-3 (4.86E-5) & 2.70E-3 (8.29E-4) & \textbf{$<$ 2.2E-16} \\
		collections-3.1   & \textbf{1.80E-4 (1.43E-5)} & 1.64E-3 (1.56E-5) & 2.25E-3 (5.11E-5) & 1.85E-3 (4.06E-4) & \textbf{$<$ 2.2E-16} \\
		collections-3.2   & \textbf{1.91E-4 (1.67E-5)} & 1.56E-3 (2.51E-5) & 2.11E-3 (6.53E-5) & 1.66E-3 (5.41E-4) & \textbf{$<$ 2.2E-16} \\
		collections-3.2.1 & \textbf{1.76E-4 (1.09E-5)} & 1.54E-3 (1.71E-5) & 2.11E-3 (7.23E-5) & 1.82E-3 (6.38E-4) & \textbf{$<$ 2.2E-16} \\\hline
		lang-3.0          & \textbf{9.67E-5 (7.90E-6)} & 1.74E-3 (2.50E-5) & 2.41E-3 (3.57E-5) & 1.93E-3 (5.54E-4) & \textbf{$<$ 2.2E-16} \\
		lang-3.0.1        & \textbf{9.68E-5 (4.41E-6)} & 1.75E-3 (1.54E-5) & 2.40E-3 (2.82E-5) & 1.80E-3 (5.08E-4) & \textbf{$<$ 2.2E-16} \\
		lang-3.1          & \textbf{1.43E-4 (1.48E-5)} & 1.84E-3 (3.51E-5) & 2.39E-3 (3.35E-5) & 1.72E-3 (4.25E-4) & \textbf{$<$ 2.2E-16} \\
		lang-3.2          & \textbf{1.78E-4 (3.33E-5)} & 1.85E-3 (3.45E-5) & 2.41E-3 (2.95E-5) & 1.84E-3 (4.46E-4) & \textbf{$<$ 2.2E-16} \\\hline
		validator-1.4.0   & \textbf{1.72E-4 (6.38E-6)} & 1.71E-3 (7.04E-5) & 3.19E-3 (8.53E-5) & 2.37E-3 (5.94E-4) & \textbf{$<$ 2.2E-16} \\
		validator-1.4.1   & \textbf{1.85E-4 (9.46E-6)} & 1.77E-3 (9.72E-5) & 3.13E-3 (1.04E-4) & 2.75E-3 (5.54E-4) & \textbf{$<$ 2.2E-16} \\
		validator-1.5.0   & \textbf{1.83E-4 (5.53E-6)} & 1.82E-3 (9.32E-5) & 3.04E-3 (8.97E-5) & 2.55E-3 (8.34E-4) & \textbf{$<$ 2.2E-16} \\
		validator-1.5.1   & \textbf{1.90E-4 (1.10E-5)} & 1.79E-3 (9.93E-5) & 3.07E-3 (8.16E-5) & 2.68E-3 (7.56E-4) & \textbf{$<$ 2.2E-16} \\\hline
		jfreechart-1.0.0  & \textbf{1.60E-4 (1.52E-6)} & 1.74E-3 (1.96E-5) & 2.72E-3 (4.59E-5) & 1.63E-3 (2.44E-4) & \textbf{$<$ 2.2E-16} \\
		jfreechart-1.0.1  & \textbf{1.50E-4 (6.29E-6)} & 1.56E-3 (3.65E-5) & 2.74E-3 (4.37E-5) & 1.54E-3 (2.98E-4) & \textbf{$<$ 2.2E-16} \\
		jfreechart-1.0.2  & \textbf{1.68E-4 (7.61E-6)} & 1.58E-3 (4.08E-5) & 2.76E-3 (3.62E-5) & 1.61E-3 (3.63E-4) & \textbf{$<$ 2.2E-16} \\
		jfreechart-1.0.3  & \textbf{1.51E-4 (3.52E-6)} & 1.53E-3 (3.32E-5) & 2.74E-3 (4.91E-5) & 1.65E-3 (4.29E-4) & \textbf{$<$ 2.2E-16} \\\hline
		jgrapht-0.9.0     & \textbf{2.23E-4 (1.70E-5)} & 1.58E-3 (5.61E-5) & 3.72E-3 (7.56E-5) & 2.43E-3 (6.82E-4) & \textbf{$<$ 2.2E-16} \\
		jgrapht-0.9.1     & \textbf{2.15E-4 (1.44E-5)} & 1.86E-3 (6.78E-5) & 4.06E-3 (8.42E-5) & 2.64E-3 (7.86E-4) & \textbf{$<$ 2.2E-16} \\
		jgrapht-0.9.2     & \textbf{2.85E-4 (1.56E-5)} & 1.72E-3 (4.53E-5) & 3.72E-3 (5.88E-5) & 2.25E-3 (5.48E-4) & \textbf{$<$ 2.2E-16} \\
		jgrapht-1.0.0     & \textbf{2.27E-4 (1.78E-5)} & 1.68E-3 (4.42E-5) & 3.65E-3 (7.08E-5) & 2.29E-3 (6.34E-4) & \textbf{$<$ 2.2E-16} \\\hline
		joda-time-2.8     & \textbf{1.26E-4 (9.44E-6)} & 1.66E-3 (2.72E-5) & 1.66E-3 (2.43E-5) & 1.25E-3 (2.44E-4) & \textbf{$<$ 2.2E-16} \\
		joda-time-2.8.1   & \textbf{1.11E-4 (5.59E-6)} & 1.70E-3 (2.36E-5) & 1.66E-3 (3.10E-5) & 1.30E-3 (3.02E-4) & \textbf{$<$ 2.2E-16} \\
		joda-time-2.8.2   & \textbf{1.13E-4 (8.08E-6)} & 1.71E-3 (2.79E-5) & 1.65E-3 (3.35E-5) & 1.33E-3 (3.46E-4) & \textbf{$<$ 2.2E-16} \\
		joda-time-2.9     & \textbf{1.25E-4 (6.08E-6)} & 1.68E-3 (2.02E-5) & 1.67E-3 (2.77E-5) & 1.36E-3 (3.47E-4) & \textbf{$<$ 2.2E-16} \\\hline
		ognl-3.1          & \textbf{9.35E-4 (4.91E-5)} & 5.87E-3 (1.12E-4) & 6.74E-3 (1.06E-4) & 5.19E-3 (9.84E-4) & \textbf{$<$ 2.2E-16} \\
		ognl-3.1.1        & \textbf{9.22E-4 (5.10E-5)} & 5.75E-3 (7.91E-5) & 6.53E-3 (8.59E-5) & 4.56E-3 (1.02E-3) & \textbf{$<$ 2.2E-16} \\
		ognl-3.1.2        & \textbf{1.06E-3 (1.12E-5)} & 5.65E-3 (1.05E-4) & 6.49E-3 (1.11E-4) & 4.51E-3 (1.08E-3) & \textbf{$<$ 2.2E-16} \\
		ognl-3.1.3        & \textbf{1.06E-3 (9.70E-6)} & 6.01E-3 (1.28E-4) & 6.82E-3 (7.98E-5) & 5.14E-3 (1.14E-3) & \textbf{$<$ 2.2E-16} \\\hline
		wire-2.0.0        & \textbf{5.08E-4 (2.26E-5)} & 3.13E-3 (3.11E-4) & 5.77E-3 (2.82E-4) & 4.13E-3 (1.02E-3) & \textbf{$<$ 2.2E-16} \\
		wire-2.0.1        & \textbf{4.68E-4 (2.34E-5)} & 3.18E-3 (2.38E-4) & 5.69E-3 (3.37E-4) & 3.99E-3 (1.07E-3) & \textbf{$<$ 2.2E-16} \\
		wire-2.0.2        & \textbf{4.61E-4 (1.48E-5)} & 3.05E-3 (1.55E-4) & 5.79E-3 (3.66E-4) & 4.09E-3 (1.19E-3) & \textbf{$<$ 2.2E-16} \\
		wire-2.0.3        & \textbf{4.68E-4 (1.83E-5)} & 3.03E-3 (3.49E-4) & 5.62E-3 (2.70E-4) & 3.76E-3 (1.02E-3) & \textbf{$<$ 2.2E-16} \\ \bottomrule
	\end{tabular}
\end{table*}

We can observe that, overall, the strategies generated by Sentinel are able to outperform the conventional ones according to both indicators. More precisely, for 70 out of 80 statistical comparisons, Sentinel strategies outperformed the conventional ones with significant differences. Only for {\it wire} SM obtained significantly better HV results, but not for IGD. For \textit{codec-1.5/1.6/1.11} and \textit{beanutils-1.8.1/1.8.2/1.8.3}, SM showed no statistical significant difference regarding HV, but fell behind in the IGD comparison.

Table~\ref{tab:Effect Size} shows the Vargha--Delaney $\hat{A}_{\mathit{12}}$ effect size results for HV and IGD. The group of strategies generated by Sentinel is subject A and each conventional strategy group -- shown in columns from two to seven -- is subject B. Large differences in favour of Sentinel strategies are highlighted in bold.
The strategies generated by Sentinel obtained large effect size in 227 out of 240 comparisons (i.e.\ 95\% of the cases), while SM and ROS obtained favourable large effect size only in one system: SM for all versions of {\it wire} and ROS for {\it wire-2.0.3}.

\begin{table}[tb]
	\centering
	\caption{\textbf{RQ2--RQ3:}  Effect size results  (L = large, M=medium, S=small, N = negligible) of Sentinel-generated strategies vs. conventional strategies. Values greater than $0.5$ are better for Sentinel (best in bold).}
	\label{tab:Effect Size}
	\fontsize{6pt}{6pt}\selectfont
	\def\arraystretch{1.2}
	\newcolumntype{R}[1]{>{\raggedleft\arraybackslash}p{#1}}
	\begin{tabular}{lrrrrrr}
		\toprule
		\multirow{2}{*}{Program} &                  \multicolumn{3}{c|}{HV}                  &             \multicolumn{3}{c}{IGD}              \\
		                         &                SM &               RMS &               \multicolumn{1}{r|}{ROS} &             SM &            RMS &            ROS \\ \midrule
		beanutils-1.8.0          & \textbf{0.78 (L)} & \textbf{0.94 (L)} & \textbf{0.84 (L)} & \textbf{0 (L)} & \textbf{0 (L)} & \textbf{0 (L)} \\
		beanutils-1.8.1          &          0.63 (S) & \textbf{0.96 (L)} & \textbf{0.80 (L)} & \textbf{0 (L)} & \textbf{0 (L)} & \textbf{0 (L)} \\
		beanutils-1.8.2          &          0.70 (M) & \textbf{0.88 (L)} & \textbf{0.77 (L)} & \textbf{0 (L)} & \textbf{0 (L)} & \textbf{0 (L)} \\
		beanutils-1.8.3          &          0.64 (S) & \textbf{0.91 (L)} & \textbf{0.80 (L)} & \textbf{0 (L)} & \textbf{0 (L)} & \textbf{0 (L)} \\\hline
		codec-1.4                & \textbf{0.88 (L)} &    \textbf{1 (L)} &    \textbf{1 (L)} & \textbf{0 (L)} & \textbf{0 (L)} & \textbf{0 (L)} \\
		codec-1.5                &          0.69 (M) &    \textbf{1 (L)} &    \textbf{1 (L)} & \textbf{0 (L)} & \textbf{0 (L)} & \textbf{0 (L)} \\
		codec-1.6                &          0.45 (N) &    \textbf{1 (L)} &    \textbf{1 (L)} & \textbf{0 (L)} & \textbf{0 (L)} & \textbf{0 (L)} \\
		codec-1.11               & \textbf{0.88 (L)} &    \textbf{1 (L)} &    \textbf{1 (L)} & \textbf{0 (L)} & \textbf{0 (L)} & \textbf{0 (L)} \\\hline
		collections-3.0          &    \textbf{1 (L)} &    \textbf{1 (L)} &    \textbf{1 (L)} & \textbf{0 (L)} & \textbf{0 (L)} & \textbf{0 (L)} \\
		collections-3.1          &    \textbf{1 (L)} &    \textbf{1 (L)} &    \textbf{1 (L)} & \textbf{0 (L)} & \textbf{0 (L)} & \textbf{0 (L)} \\
		collections-3.2          &    \textbf{1 (L)} &    \textbf{1 (L)} &    \textbf{1 (L)} & \textbf{0 (L)} & \textbf{0 (L)} & \textbf{0 (L)} \\
		collections-3.2.1        &    \textbf{1 (L)} &    \textbf{1 (L)} &    \textbf{1 (L)} & \textbf{0 (L)} & \textbf{0 (L)} & \textbf{0 (L)} \\\hline
		lang-3.0                 &    \textbf{1 (L)} & \textbf{0.99 (L)} &    \textbf{1 (L)} & \textbf{0 (L)} & \textbf{0 (L)} & \textbf{0 (L)} \\
		lang-3.0.1               &    \textbf{1 (L)} &    \textbf{1 (L)} &    \textbf{1 (L)} & \textbf{0 (L)} & \textbf{0 (L)} & \textbf{0 (L)} \\
		lang-3.1                 &    \textbf{1 (L)} &    \textbf{1 (L)} &    \textbf{1 (L)} & \textbf{0 (L)} & \textbf{0 (L)} & \textbf{0 (L)} \\
		lang-3.2                 &    \textbf{1 (L)} &    \textbf{1 (L)} &    \textbf{1 (L)} & \textbf{0 (L)} & \textbf{0 (L)} & \textbf{0 (L)} \\\hline
		validator-1.4.0          &    \textbf{1 (L)} &    \textbf{1 (L)} &    \textbf{1 (L)} & \textbf{0 (L)} & \textbf{0 (L)} & \textbf{0 (L)} \\
		validator-1.4.1          &    \textbf{1 (L)} &    \textbf{1 (L)} &    \textbf{1 (L)} & \textbf{0 (L)} & \textbf{0 (L)} & \textbf{0 (L)} \\
		validator-1.5.0          &    \textbf{1 (L)} &    \textbf{1 (L)} &    \textbf{1 (L)} & \textbf{0 (L)} & \textbf{0 (L)} & \textbf{0 (L)} \\
		validator-1.5.1          &    \textbf{1 (L)} &    \textbf{1 (L)} &    \textbf{1 (L)} & \textbf{0 (L)} & \textbf{0 (L)} & \textbf{0 (L)} \\\hline
		jfreechart-1.0.0         &    \textbf{1 (L)} &    \textbf{1 (L)} &    \textbf{1 (L)} & \textbf{0 (L)} & \textbf{0 (L)} & \textbf{0 (L)} \\
		jfreechart-1.0.1         &    \textbf{1 (L)} &    \textbf{1 (L)} &    \textbf{1 (L)} & \textbf{0 (L)} & \textbf{0 (L)} & \textbf{0 (L)} \\
		jfreechart-1.0.2         &    \textbf{1 (L)} &    \textbf{1 (L)} &    \textbf{1 (L)} & \textbf{0 (L)} & \textbf{0 (L)} & \textbf{0 (L)} \\
		jfreechart-1.0.3         &    \textbf{1 (L)} &    \textbf{1 (L)} &    \textbf{1 (L)} & \textbf{0 (L)} & \textbf{0 (L)} & \textbf{0 (L)} \\\hline
		jgrapht-0.9.0            & \textbf{0.99 (L)} &    \textbf{1 (L)} &    \textbf{1 (L)} & \textbf{0 (L)} & \textbf{0 (L)} & \textbf{0 (L)} \\
		jgrapht-0.9.1            & \textbf{0.97 (L)} &    \textbf{1 (L)} &    \textbf{1 (L)} & \textbf{0 (L)} & \textbf{0 (L)} & \textbf{0 (L)} \\
		jgrapht-0.9.2            & \textbf{0.95 (L)} &    \textbf{1 (L)} &    \textbf{1 (L)} & \textbf{0 (L)} & \textbf{0 (L)} & \textbf{0 (L)} \\
		jgrapht-1.0.0            & \textbf{0.92 (L)} &    \textbf{1 (L)} &    \textbf{1 (L)} & \textbf{0 (L)} & \textbf{0 (L)} & \textbf{0 (L)} \\\hline
		joda-time-2.8            &    \textbf{1 (L)} &    \textbf{1 (L)} &    \textbf{1 (L)} & \textbf{0 (L)} & \textbf{0 (L)} & \textbf{0 (L)} \\
		joda-time-2.8.1          &    \textbf{1 (L)} &    \textbf{1 (L)} &    \textbf{1 (L)} & \textbf{0 (L)} & \textbf{0 (L)} & \textbf{0 (L)} \\
		joda-time-2.8.2          &    \textbf{1 (L)} &    \textbf{1 (L)} &    \textbf{1 (L)} & \textbf{0 (L)} & \textbf{0 (L)} & \textbf{0 (L)} \\
		joda-time-2.9            &    \textbf{1 (L)} &    \textbf{1 (L)} &    \textbf{1 (L)} & \textbf{0 (L)} & \textbf{0 (L)} & \textbf{0 (L)} \\\hline
		ognl-3.1                 & \textbf{0.99 (L)} &    \textbf{1 (L)} &    \textbf{1 (L)} & \textbf{0 (L)} & \textbf{0 (L)} & \textbf{0 (L)} \\
		ognl-3.1.1               &    \textbf{1 (L)} &    \textbf{1 (L)} &    \textbf{1 (L)} & \textbf{0 (L)} & \textbf{0 (L)} & \textbf{0 (L)} \\
		ognl-3.1.2               &    \textbf{1 (L)} &    \textbf{1 (L)} &    \textbf{1 (L)} & \textbf{0 (L)} & \textbf{0 (L)} & \textbf{0 (L)} \\
		ognl-3.1.3               &    \textbf{1 (L)} &    \textbf{1 (L)} &    \textbf{1 (L)} & \textbf{0 (L)} & \textbf{0 (L)} & \textbf{0 (L)} \\\hline
		wire-2.0.0               &          0.03 (L) &    \textbf{1 (L)} &          0.57 (N) & \textbf{0 (L)} & \textbf{0 (L)} & \textbf{0 (L)} \\
		wire-2.0.1               &          0.04 (L) &    \textbf{1 (L)} &          0.40 (S) & \textbf{0 (L)} & \textbf{0 (L)} & \textbf{0 (L)} \\
		wire-2.0.2               &          0.05 (L) &    \textbf{1 (L)} &          0.45 (N) & \textbf{0 (L)} & \textbf{0 (L)} & \textbf{0 (L)} \\
		wire-2.0.3               &             0 (L) &    \textbf{1 (L)} &          0.19 (L) & \textbf{0 (L)} & \textbf{0 (L)} & \textbf{0 (L)} \\ \bottomrule
	\end{tabular}
\end{table}

Therefore, we can conclude that the strategies generated by Sentinel perform statistically significantly better than the conventional strategies for the majority of the systems tested (i.e.\ 95\% of the cases). This positively answers our research question:

~

\noindent
\fbox{\begin{minipage}{0.98 \columnwidth}
\textbf{RQ2}: \textit{The strategies generated by Sentinel present a significantly better trade-off between execution time and score approaching than the conventional strategies RMS, ROS and SM.}
\end{minipage}
}

\subsection{RQ3 -- Sentinel Strategies Effectiveness Over Time}
\label{sec:Empirical Study:Results:RQ3}

Training any automated or manual approach to solve a problem comes with an additional cost. Such a cost is alleviated if the trained model (in our case, the strategies found) can be re-used over time to solve subsequent unseen instances of the problem. In this section, we discuss and compare the training time of Sentinel and conventional strategies, and then assess if the strategies learnt at a certain time (i.e.\ for a given version of a system) can be re-used later on for subsequent versions without the need of generating new ones (and hence to re-train Sentinel incurring in additional training costs).

Table~\ref{tab:Training Times} shows the average training time for Sentinel and conventional strategies (i.e.\ the cost of manually comparing all RMS, ROS and SM strategies in order to choose the best one). We can observe that, for six out of 10 systems, the training cost of Sentinel is lower than experimenting and comparing RMS, ROS and SM due to the caching system used\footnote{The caching mechanism is described in Section~\ref{sec:Sentinel:Implementation Aspects}. For some of the systems caching is faster than executing all conventional strategies, but for others this incurs in a costlier overhead for setting up one JVM instance for each mutant.}, yet the training cost of all these approaches is not negligible.  

\begin{table}[tb]
	\centering
	\caption{\textbf{RQ3:} Average training time of Sentinel and conventional strategies. Time format ``hh:mm:ss''.}
	\label{tab:Training Times}
	\fontsize{8pt}{8pt}\selectfont
	\def\arraystretch{1.2}
	\begin{tabular}{lrr}
		\toprule
		Program     & Sentinel & Conventional Strategies \\ \midrule
		beanutils   &          04:12:23 &       01:36:24 \\
		codec       &          01:22:11 &       01:56:23 \\
		collections &          07:24:22 &       02:37:29 \\
		lang        &          11:23:49 &       04:23:09 \\
		validator   &             38:44 &       01:06:21 \\
		jfreechart  &          07:51:33 &       11:38:16 \\
		jgrapht     &          01:49:51 &       04:41:22 \\
		joda-time   &          03:25:33 &       06:54:45 \\
		ognl        &          01:19:44 &          51:27 \\
		wire        &             01:41 &          11:58 \\ \bottomrule
	\end{tabular}
\end{table}

If we had to train for each version of a software, the training cost would make their use ineffective. Fortunately, the results reported in Section~\ref{sec:Empirical Study:Results:RQ2} show that reusing strategies over time is possible without significant loss in their effectiveness. As we can observe from Tables~\ref{tab:HV Results}-\ref{tab:Effect Size}, not only the strategies generated by Sentinel are able to hold their good results across multiple versions, but they also outperform conventional strategies. For seven out of the 10 analysed software programs, the strategies generated by Sentinel showed favourable large statistical differences for all the subsequent versions considering both IGD and HV. For two of those 10 programs, the strategies generated by Sentinel showed no statistical differences to SM strategies for HV in subsequent versions, and outperformed ROS and RMS with large statistical differences. Looking at the IGD results, Sentinel strategies were able to maintain large statistical differences throughout all the evaluated versions for all the 10 systems. The systems \textit{beanutils} and \textit{codec} are the only two for which Sentinel's generated strategies did not maintain their superiority considering HV values for subsequent versions. The code churn of \textit{beanutils} is one of the smallest between the programs (Table~\ref{tab:Programs}), whereas \textit{codec}'s churn is somewhat close to the median. For systems in between those churn sizes (or in the extremities for that matter), Sentinel's strategies statistically outperformed the conventional ones in subsequent versions. Hence, we did not observe any relation between code churn and effectiveness degradation of the trained strategies.

Summarising, for 95\% of comparisons, Sentinel strategies learnt on a previous version can be reused as they are statistically better or equivalent to conventional ones when used for subsequent versions of a same software. Therefore, we can state that, overall, Sentinel strategies can be trained on a given system's version and then reused up to three subsequent versions without either affecting their mutant reduction effectiveness, or incurring in additional training costs. This answers positively our third and last research question:

~

\noindent
\fbox{\begin{minipage}{0.98 \columnwidth}
		\textbf{RQ3}: \textit{The strategies generated by Sentinel can be effectively used for subsequent versions of a software.}
	\end{minipage}
}

\subsection{Discussion}
\label{sec:Empirical Study:Results:Discussion}

The results presented in this section reveal that arbitrarily selecting and configuring a strategy may lead to unsatisfactory time reduction, mutation score degradation or, in the worst case, both. For instance, \textit{joda-time-2.8} has approximately 10,000 mutants, thus RMS 10\% gives us 1,000 mutants. The average result for RMS 10\% (lower cluster of black squares in Figure~\ref{fig:Pareto}) is approximately 15\% time cost and a mutation score approaching of only 55\% on average. For \textit{jfreechart-1.0.0}, there are approximately 20,000 mutants. The lower cluster of black squares of Figure~\ref{fig:Pareto}(a) represents RMS 10\%, which selects 2,000 mutants. The average result for this strategy is around 5\% CPU time cost but approaching only 75\% of the original mutation score. In other words, the same percentage of mutants in different systems yields different results in both mutation score and CPU time: while 1,000 mutants in \textit{joda-time-2.8} yields 15\% of time and 55\% of score approaching, 2,000 mutants in \textit{jfreechart-1.0.0} yields only 5\% of time and 75\% score. We also observed such a discrepancy for the other systems. With these examples in mind and based on what we observed during our experiments, we can state that choosing the right parameters and strategies is crucial for effectively reducing the mutant set and that mutant reduction strategies should be carefully tuned for each software system. 

These results show that there is room for improvement of mutant reduction strategies and doing it automatically seems preferable in most cases. As far as we could observe, the strategies generated by Sentinel obtain the best trade-off between mutation score approaching and execution time, and generally maintain this best performance over subsequent software versions. Sentinel also removes from the hands of the engineer the tedious, error-prone and costly task of selecting and configuring strategies. We advocate that Sentinel should be used in two situations where the mutant reduction is more impactful: i)~when the time taken to execute all mutants is impracticable; and ii)~when the mutation testing is performed several times during the software development process. Indeed, how long is ``impracticable time'' and how many is ``several times'' depend on many factors such as environmental constraints, software testing budget and even on the required reliability level involved in the testing process. In any case, if the testers find themselves in at least one of those two situations, then the training cost of Sentinel can be a good price to pay for the significant mutant reduction trade-off provided by the generated strategies, specially considering that doing an exhaustive manual experimentation to find the best conventional strategy was more expensive than Sentinel training for six out of 10 systems (as seen in Section~\ref{sec:Empirical Study:Results:RQ3}). Moreover, Sentinel's training cost can be further reduced by running it in parallel~\cite{Sarro2020GECCO,GeronimoFMS12,FerrucciSalzaSarro,DiMartino2013,FerrucciSKS15,SalzaFS16,SalzaFS16a}.

Finally, based on our data, we observed that number of mutants is not always an accurate surrogate for cost. Thus, we advise engineers to measure execution time when comparing the cost of mutation testing for a better assessment~\cite{Guizzo2020}.

\section{Related Work}
\label{sec:Related Work}

In this section we discuss previous work on hyper-heuristics for mutation testing and on mutant reduction strategies. Comprehensive surveys on mutation testing can be found elsewhere~\cite{Jia2011, Bashir2012, Papadakis2017}. We also refer the reader to a recent systematic review on mutation cost reduction by Pizzoleto et al.~\cite{Pizzoleto2019} for a comprehensive description of cost reduction strategies used in the literature.

\subsection{Hyper-Heuristics for Mutation Testing}
\label{sec:Related Work:Hyper-Heuristics in Mutation Testing}

The use of hyper-heuristics for Software Engineering have been recently investigated providing promising results for project management~\cite{Sarro2017}, software clustering~\cite{Kumari2016}, combinatorial interaction testing~\cite{Jia2015}, integration and test order~\cite{Guizzo2015,  Mariani2016, Guizzo2017} and feature models testing~\cite{Strickler2016}. Few studies have investigated hyper-heuristics for mutation testing~\cite{Ferreira2017, Strickler2016, Filho2017, Lima2017}. The work of Ferreira et al.~\cite{Ferreira2017} and Strickler et al.~\cite{Strickler2016} use the online Hyper-heuristic for the Integration and Test Order Problem (HITO)~\cite{Guizzo2015}. The idea is to automatically select mutation and crossover operators of MOEAs. Such hyper-heuristic is applied in a different problem than the one for which it was originally proposed~\cite{Guizzo2015}: the selection of products for testing Software Product Lines (SPLs). HITO achieved competitive results with respect to conventional MOEAs. Jakuboviski Filho et al.~\cite{Filho2017} also applied an offline hyper-heuristic based on GE (proposed by Mariani et al.~\cite{Mariani2016}) to automatically generate MOEAs for the same SPL problem. The generated MOEAs were compared to NSGA--II and to HITO~\cite{Guizzo2015} and showed competitive results with respect to HITO and generally better results than NSGA--II.

Such studies~\cite{Ferreira2017, Strickler2016, Filho2017} have as main goal the selection of test data (products) to satisfy the mutation testing of Feature Models. Sentinel has a distinct goal, since it is used to generate strategies to reduce the set of mutants. While Sentinel is an off-line generation hyper-heuristic for mutant reduction strategies, the hyper-heuristics used in the related work are on-line selection hyper-heuristics for adaptively selecting evolutionary operators~\cite{Ferreira2017, Strickler2016} or off-line generation of MOEAs~\cite{Filho2017}.
Lima and Vergilio~\cite{Lima2017} proposed a hyper-heuristic to automatically select crossover and mutation operators of a MOEA in order to generate HOMs. The goal of this work is different from ours, since the approach searches to find the best HOMs by optimising three objectives: i)~minimising the total number of generated HOMs; ii)~maximising the number of revealed subtle faults (i.e.\ faults that can only be revealed by the HOM); and iii)~maximising the number of strongly subsuming HOMs.

\subsection{Mutant Reduction Strategies}
\label{sec:Related Work:Mutants Reduction Strategies}

Mathur and Wong~\cite{Mathur1994} applied mutant sampling on C programs and were able to reduce the number of mutants by 90\% while losing only $0.16$ points in the mutation score. Wong et al.~\cite{Wong1995} used a similar strategy for the same testing scenario, by selecting only a percentage of mutants from each available mutation operator and showed that this strategy was equally efficient. Papadakis and Malevris~\cite{Papadakis2010} conducted a study on C programs by sampling 10\%, 20\%, 30\%, 40\%, 50\% and 60\% mutants and found that the mutation effectiveness is reduced by 26\%--6\% using such a strategy. Sahinoglu and Spafford~\cite{Sahinoglu1990} proposed a strategy based on the Bayesian Inference for sampling mutants, which was evaluated for the unit testing of C programs showing that a better mutant reduction was obtained with respect to a conventional strategy.

Hussain et al.~\cite{Hussain2008} proposed mutant clustering strategies which use clustering algorithms to select mutants from different clusters, because test cases that kill a mutant likely kill the mutants in the same cluster. The problem is that all the test cases must be executed against all mutants before applying the clustering algorithm, which is a very costly task. Another problem is how to define the number of centroids. Ji et al.~\cite{Ji2009} propose a clustering strategy that uses information about the program domain to perform the mutant clustering. The advantage of this strategy is the generation and selection of mutants before executing the test cases. The authors evaluated this strategy in the unit test of a single Java program and observed that the strategy is capable of maintaining the mutation score as high as $0.9$ while reducing in 50\% the number of mutants.

Mathur~\cite{Mathur1991} omitted the two most costly operators out of the 22 mutation operators available in the Mothra tool~\cite{King1991} used to mutate Fortran 77 programs. These two operators generate between 30\% and 40\% of all mutants. Results show a mutation score of $0.9999$ and a mutant reduction of 24\%. Offutt et al.~\cite{Offutt1993} extended the work of Mathur~\cite{Mathur1991} by omitting the four and six more costly operators. For the four operators exclusion, the mutant reduction was approximately 40\% with a $0.9984$ score, whereas for the six operator exclusion the obtained score was $0.8871$ with a 60\% reduction. Delamaro et al.~\cite{Delamaro2014} performed an empirical study using only one mutation operator for C programs. The hypothesis is that using only one powerful operator should be enough to perform the mutation testing. The mutation operator responsible for removing lines of code is the most efficient one, since it generates approximately 3.26\% of all mutants and obtains a score of approximately $0.92$.

Barbosa et al.~\cite{Barbosa2001} defined some guidelines for the selection of essential operators (a set with the best operators). The authors obtained a set of 10 essential operators among the 77 available in the Proteum tool~\cite{Delamaro2001}. This set was able to reduce by 65\% the number of mutants while maintaining a mutation score of $0.996$. Vincenzi et al.~\cite{Vincenzi1999} also used guidelines like these, but for the mutation in the integration testing phase (interface mutation). The results showed that it was possible to reduce by 73\% the number of generated mutants while keeping the mutation score as high as $0.998$. Namin et al.~\cite{Namin2008} proposed a statistical analysis procedure with a linear regression model to find essential operators. A set of 28 operators was found which generated only 8\% of all mutants and obtained an acceptable mutation score.

Previous work also compared different conventional strategies. Gopinath et al.~\cite{Gopinath2015} compared mutant sampling strategies that sample a constant number of mutants to strategies that sample a percentage, finding that strategies that use a constant number of mutants (as low as 1,000) can obtain an accurate mutation score (approximately 93\%) when compared to the whole set of mutants. Zhang et al.~\cite{Zhang2013} discovered that using 5\% random mutant sampling in combination with operator-based selection strategies can greatly reduce the number of mutants while maintaining a great mutation accuracy. Zhang et al.~\cite{Zhang2010} compared strategies that select mutation operators with mutant sampling strategies, showing that selecting operators is not superior to sampling random mutants. Gopinath et al.~\cite{Gopinath2016} observed similar results with a theoretical and empirical analysis and also concluded that the reduction limit over random mutant sampling strategies is 13\%. Lima et al.~\cite{Lima2016} compared SM, RMS, HOM generation strategies and an Evolutionary Algorithm for selecting mutants and test cases to reduce the mutation cost of C programs. The authors discovered that SM performed better than the other strategies and that the HOM strategies generated more mutants with a lower mutation score. 

These works diverge on the conclusion about which strategy is the best, as shown in a recent literature review~\cite{Pizzoleto2019}. In fact, it depends on the context of the test, such as program under test, testing phase and programming language. Hence, using an automatic hyper-heuristic such as Sentinel can provide good strategies that are tailored for a specific scenario. Moreover, we observed that some mutants take longer to execute than others, e.g.\ mutants that die due to time out or mutants with a fault in memory allocation. By assessing the CPU execution time we can accurately measure the cost reduction of strategies, as opposed to using solely the number of mutants~\cite{Guizzo2020}.

Recently, Petrovic and Ivankovic~\cite{Petrovic2018} reported how mutation testing is used at Google at scale. Their approach deviates from the conventional mutation analysis by changing the way mutants are generated and presented to the developer. Their tool only generates one possible mutant per covered line. Traditional mutation tools usually also only apply mutations to covered lines, however they usually do not restrict the generation of mutants to one per line. As the authors state themselves, they do not use mutation score to measure the test effectiveness, since the mutation score relies on the whole set of possible mutants and they in fact only generate a limited set of mutants using a probabilistic technique. The second difference is that the focus of their approach is to maximise the ``usefulness'' of mutants by showing only interesting mutants to the developers during code review. If the developer judges an alive mutant as irrelevant, even if killable, the mutant is simply ignored. According to the developer's feedback, a heuristic is applied to prune the ``uninteresting'' lines (dubbed ``arid lines'') from the mutation process. Therefore, the mutant set depends on historical data obtained from the developers and on the heuristic to determine which mutants are uninteresting. In conventional mutation, all mutants are considered in order to improve the mutation score.

The third difference is on the scope of the mutation. While traditional mutation usually treats a program as a whole, their approach narrows down the mutation to diffs and commits. Those differences unveil a very interesting relation between current academic literature on mutation testing and the practice in industry. The challenges faced in each scenario might be different and should be considered in future work.

\section{Conclusion}
\label{sec:Conclusion}

This paper presented Sentinel, a hyper-heuristic based on GE to generate mutant reduction strategies. Our empirical results show that Sentinel performs statistically significantly better than a Random hyper-heuristic for all of the 10 used real-world systems with large effect size in 95\% of the comparisons. Moreover, when compared to state-of-the-art strategies, the strategies generated by Sentinel achieved statistically significantly better results for 70 out of 80 comparisons (88\%), and large effect size in 95\% of the cases. Also, we investigated if the mutation strategies generated with Sentinel on a given version can be used in subsequent versions of a same system revealing positive results in 95\% of the cases. These results show that Sentinel strategies are effective to the mutant reduction problem when considering both global mutation score approaching and CPU time minimisation.

\begin{table}[]
	\centering
	\caption{Average CPU time and approaching score per strategy over 30 independent runs (CPU time is given as the percentage of time used by a strategy to perform the mutation in relation to generating and executing all mutants).}
	\fontsize{6pt}{6pt}\selectfont
	\def\arraystretch{1.2}
	\label{tab:TimeandScore}
	\begin{tabular}{lrrrrrrrr}
		\toprule
		\multirow{2}{*}{Program} &          \multicolumn{4}{c|}{Score}           & \multicolumn{4}{c}{Time}  \\
		& Sen. &   SM &  RMS & \multicolumn{1}{c|}{ROS} & Sen. &   SM &  RMS &  ROS \\ \midrule
		beanutils-1.8.0          & 0.63 & 0.48 & 0.89 &                     0.82 & 0.43 & 0.27 & 0.54 & 0.49 \\
		beanutils-1.8.1          & 0.63 & 0.48 & 0.89 &                     0.81 & 0.46 & 0.28 & 0.56 & 0.50 \\
		beanutils-1.8.2          & 0.63 & 0.48 & 0.88 &                     0.82 & 0.46 & 0.28 & 0.56 & 0.52 \\
		beanutils-1.8.3          & 0.63 & 0.48 & 0.88 &                     0.82 & 0.46 & 0.28 & 0.55 & 0.50 \\ \hline
		codec-1.4                & 0.68 & 0.80 & 0.93 &                     0.90 & 0.34 & 0.27 & 0.48 & 0.41 \\
		codec-1.5                & 0.65 & 0.83 & 0.93 &                     0.90 & 0.31 & 0.26 & 0.47 & 0.40 \\
		codec-1.6                & 0.64 & 0.82 & 0.93 &                     0.90 & 0.32 & 0.27 & 0.48 & 0.42 \\
		codec-1.11               & 0.67 & 0.81 & 0.93 &                     0.89 & 0.25 & 0.24 & 0.43 & 0.39 \\ \hline
		collections-3.0          & 0.55 & 0.56 & 0.85 &                     0.81 & 0.16 & 0.14 & 0.33 & 0.35 \\
		collections-3.1          & 0.54 & 0.54 & 0.85 &                     0.77 & 0.16 & 0.14 & 0.33 & 0.32 \\
		collections-3.2          & 0.56 & 0.53 & 0.84 &                     0.76 & 0.17 & 0.14 & 0.34 & 0.33 \\
		collections-3.2.1        & 0.56 & 0.53 & 0.84 &                     0.76 & 0.16 & 0.14 & 0.33 & 0.32 \\ \hline
		lang-3.0                 & 0.59 & 0.69 & 0.90 &                     0.83 & 0.13 & 0.12 & 0.31 & 0.29 \\
		lang-3.0.1               & 0.59 & 0.69 & 0.90 &                     0.83 & 0.12 & 0.12 & 0.29 & 0.29 \\
		lang-3.1                 & 0.59 & 0.68 & 0.90 &                     0.82 & 0.15 & 0.15 & 0.38 & 0.35 \\
		lang-3.2                 & 0.59 & 0.69 & 0.90 &                     0.82 & 0.15 & 0.16 & 0.36 & 0.35 \\ \hline
		validator-1.4.0          & 0.63 & 0.59 & 0.93 &                     0.84 & 0.35 & 0.28 & 0.52 & 0.48 \\
		validator-1.4.1          & 0.64 & 0.61 & 0.93 &                     0.86 & 0.35 & 0.28 & 0.53 & 0.50 \\
		validator-1.5.0          & 0.63 & 0.60 & 0.92 &                     0.84 & 0.36 & 0.30 & 0.56 & 0.51 \\
		validator-1.5.1          & 0.62 & 0.60 & 0.92 &                     0.85 & 0.36 & 0.30 & 0.55 & 0.51 \\ \hline
		jfreechart-1.0.0         & 0.69 & 0.59 & 0.94 &                     0.80 & 0.13 & 0.16 & 0.34 & 0.33 \\
		jfreechart-1.0.1         & 0.69 & 0.60 & 0.94 &                     0.79 & 0.11 & 0.13 & 0.27 & 0.25 \\
		jfreechart-1.0.2         & 0.69 & 0.61 & 0.94 &                     0.81 & 0.11 & 0.13 & 0.28 & 0.27 \\
		jfreechart-1.0.3         & 0.68 & 0.60 & 0.94 &                     0.81 & 0.11 & 0.13 & 0.27 & 0.26 \\ \hline
		jgrapht-0.9.0            & 0.68 & 0.74 & 0.96 &                     0.89 & 0.25 & 0.25 & 0.45 & 0.42 \\
		jgrapht-0.9.1            & 0.65 & 0.74 & 0.96 &                     0.88 & 0.26 & 0.26 & 0.46 & 0.42 \\
		jgrapht-0.9.2            & 0.68 & 0.75 & 0.96 &                     0.88 & 0.26 & 0.26 & 0.49 & 0.45 \\
		jgrapht-1.0.0            & 0.66 & 0.74 & 0.95 &                     0.87 & 0.25 & 0.25 & 0.46 & 0.42 \\ \hline
		joda-time-2.8            & 0.65 & 0.50 & 0.86 &                     0.76 & 0.26 & 0.16 & 0.39 & 0.35 \\
		joda-time-2.8.1          & 0.65 & 0.50 & 0.86 &                     0.76 & 0.28 & 0.18 & 0.42 & 0.36 \\
		joda-time-2.8.2          & 0.65 & 0.50 & 0.86 &                     0.76 & 0.28 & 0.18 & 0.41 & 0.37 \\
		joda-time-2.9            & 0.64 & 0.51 & 0.86 &                     0.76 & 0.27 & 0.18 & 0.40 & 0.35 \\ \hline
		ognl-3.1                 & 0.60 & 0.94 & 0.99 &                     0.95 & 0.05 & 0.11 & 0.33 & 0.31 \\
		ognl-3.1.1               & 0.61 & 0.93 & 0.99 &                     0.94 & 0.04 & 0.11 & 0.32 & 0.29 \\
		ognl-3.1.2               & 0.61 & 0.93 & 0.99 &                     0.94 & 0.05 & 0.13 & 0.38 & 0.34 \\
		ognl-3.1.3               & 0.59 & 0.93 & 0.99 &                     0.95 & 0.05 & 0.13 & 0.38 & 0.36 \\ \hline
		wire-2.0.0               & 0.45 & 0.81 & 0.95 &                     0.90 & 0.48 & 0.51 & 0.71 & 0.61 \\
		wire-2.0.1               & 0.42 & 0.81 & 0.95 &                     0.91 & 0.50 & 0.52 & 0.72 & 0.62 \\
		wire-2.0.2               & 0.42 & 0.80 & 0.95 &                     0.91 & 0.50 & 0.52 & 0.71 & 0.62 \\
		wire-2.0.3               & 0.42 & 0.79 & 0.95 &                     0.90 & 0.49 & 0.49 & 0.70 & 0.61 \\ \bottomrule
	\end{tabular}
\end{table}

As future work, Sentinel could be deployed as a service in the cloud in order to further reduce its training costs and facilitate its adoption in practice. Further scientific aspects of our proposal could  be investigated by experimenting Sentinel with other fitness functions and by assessing whether the generated strategies can be used for cross-projects mutation testing, i.e.\ if strategies generated using a given program as training subject can obtain good results when used for mutation testing of other programs from similar domains. There are other cost reduction techniques that do not rely on mutant reduction such as parallelism, weak mutation, predictive mutation~\cite{Zhang2016}, the approach proposed by Petrovic and Ivankovic~\cite{Petrovic2018}, and test case selection techniques such as Regression Mutation Testing (RMT)~\cite{Chen2018}. Such techniques can be used in combination with mutant reduction strategies and in different phases of the testing activity. For example, RMT can be applied during regression testing, and then the mutant reduction strategies generated by Sentinel can be used to further reduce the mutant set. Their interoperability should be investigated in the future.

We have made a Java implementation of Sentinel (together with our data and results) publicly available~\cite{SentinelWebsite,SentinelGH} to allow for replication and extension of our study, and to facilitate Sentinel's adoption by both practitioners and researchers, who can tackle this rich avenue of future work.

\section*{Acknowledgements}
This work is supported by the Microsoft Azure Research Award (MS-AZR-0036P), the Brazilian funding agencies CAPES and CNPq (grants 307762/2015-7 and 305968/2018), and the ERC advanced fellowship grant no. 741278 (EPIC).

%\vspace{-1em}
\bibliographystyle{IEEEtran}
\bibliography{bib}

\begin{IEEEbiography}[{\includegraphics[width=1in,height=1.25in,clip,keepaspectratio]{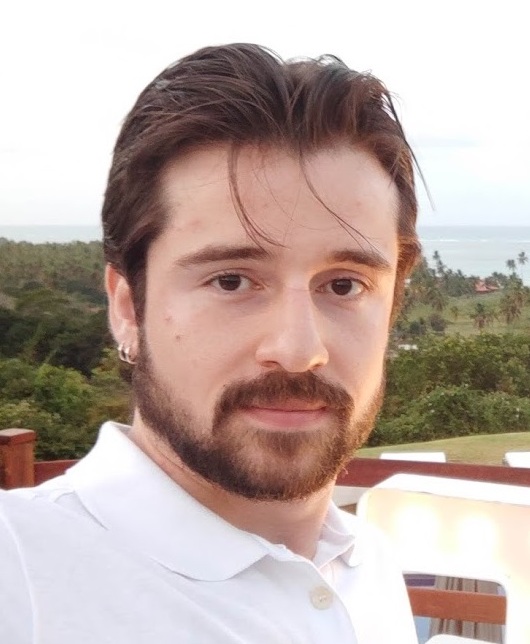}}]{Giovani Guizzo} is a Research Fellow of CREST (Centre for Research on Evolution, Search, and Testing) at University College London. Giovani is currently doing research on Search Based Software Engineering (SBSE). He has worked mainly with Search Based Software Design, Search Based Software Testing, Mutation Testing, Hyper-Heuristics and Multi-Objective Optimisation.
\end{IEEEbiography}

\begin{IEEEbiography}[{\includegraphics[width=1in,height=1.25in,clip,keepaspectratio]{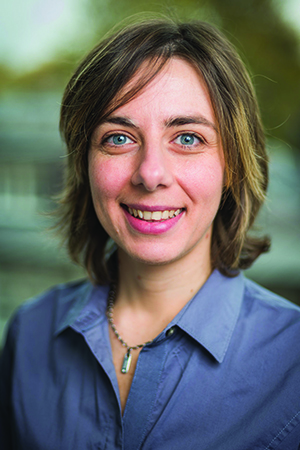}}]{Federica Sarro} is an Associate Professor at University College London. 
Her research covers Software Analytics, Empirical Software Engineering and Search-Based Software Engineering, applied to software project management, software sizing, software testing, and app store analysis. 
On these topics she has published more than 70 peer-reviewed articles and received several international awards.
She has also served on many steering, organisation and programme committees, and editorial boards of well-renowned venues such as ICSE, FSE, TSE, TOSEM and EMSE.
\end{IEEEbiography}

\begin{IEEEbiography}[{\includegraphics[width=1in,height=1.25in,clip,keepaspectratio]{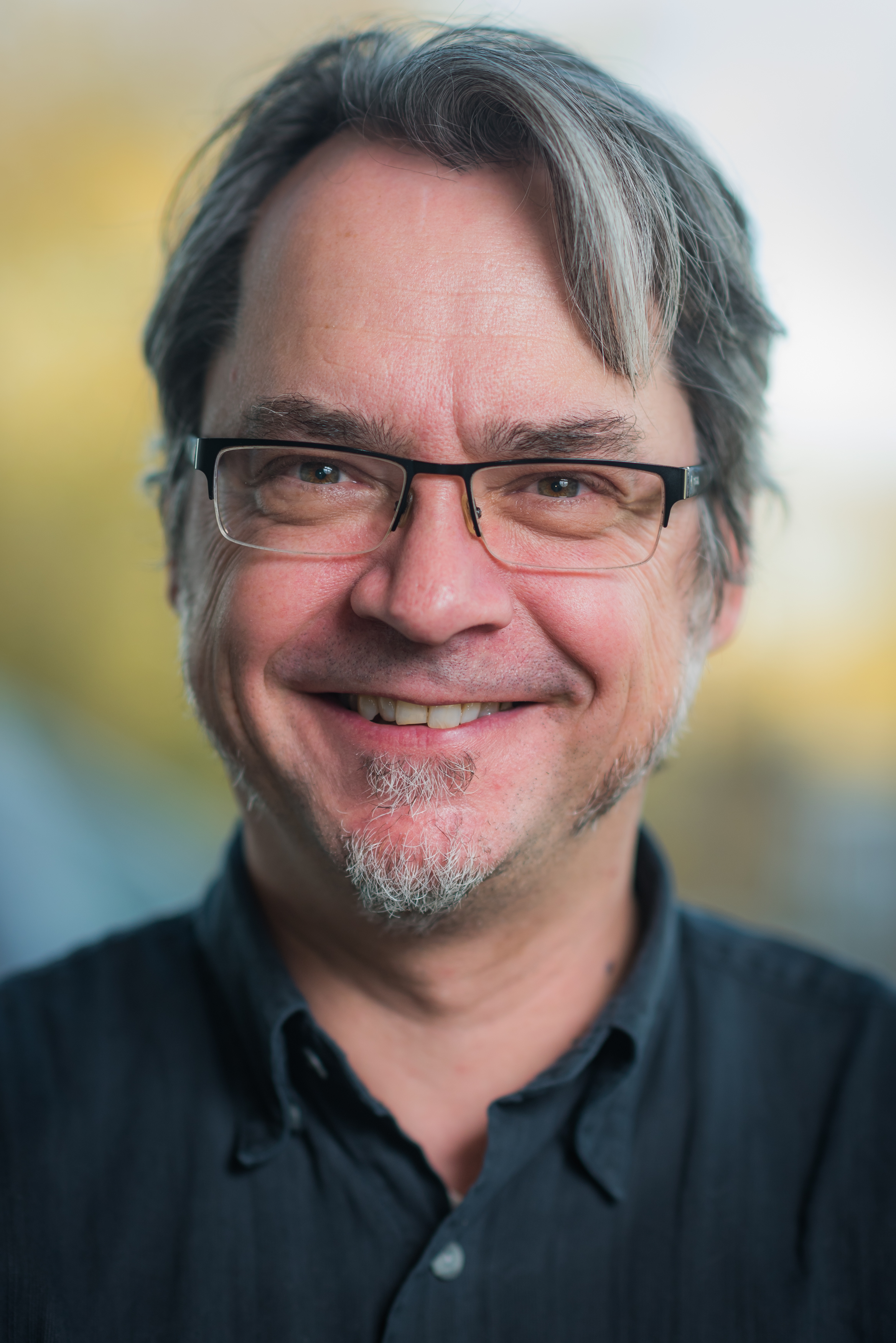}}]{Jens Krinke} is an Associate Professor in the Software Systems Engineering Group at the University College London, where he is Director of CREST, the Centre for Research on Evolution, Search, and Testing. His main focus is software analysis for software engineering purposes. His current research interests include software similarity, modern code review, and mutation testing. He is well known for his work on program slicing and clone detection.
\end{IEEEbiography}

\begin{IEEEbiography}[{\includegraphics[width=1in,height=1.25in,clip,keepaspectratio]{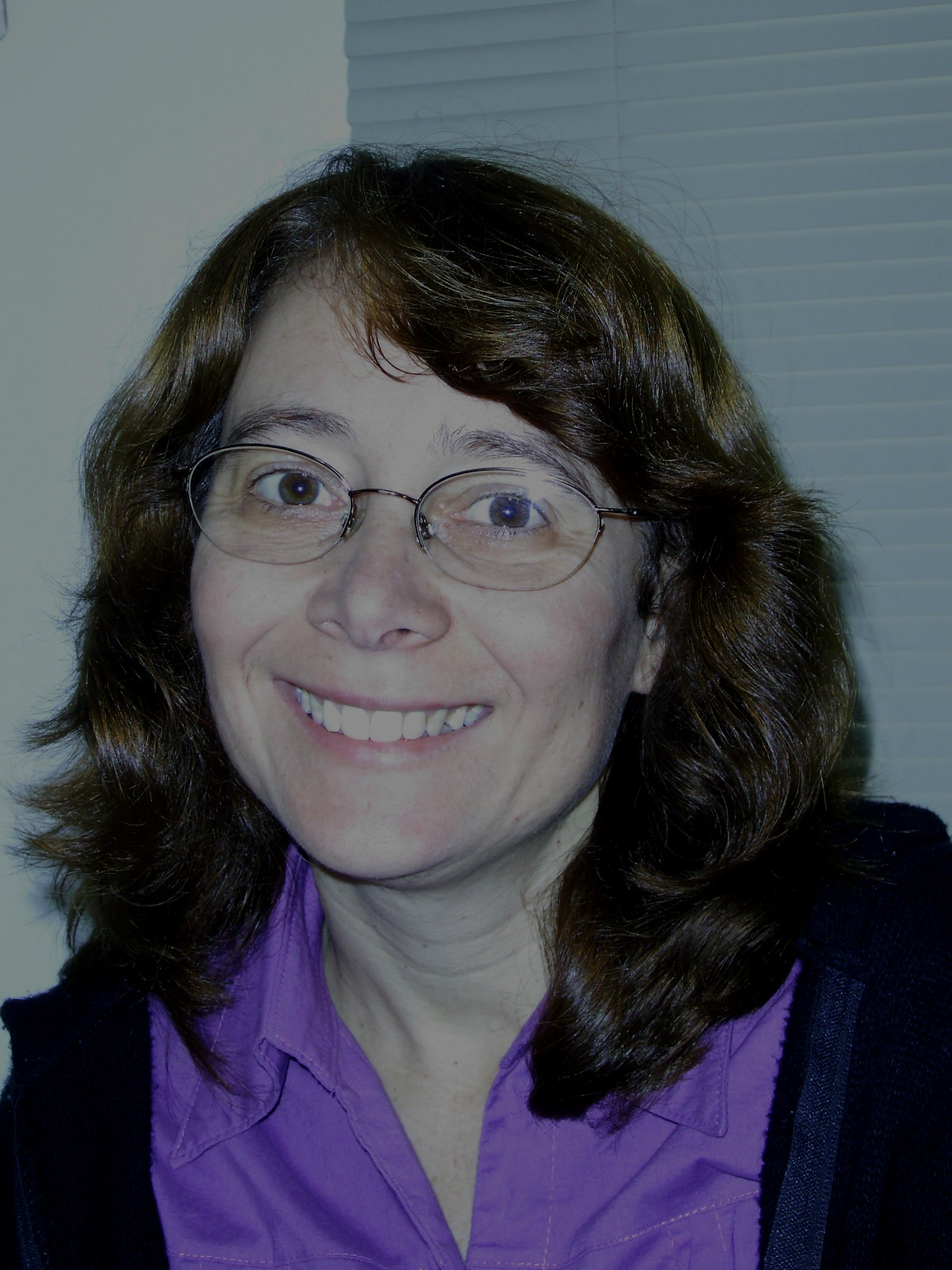}}]{Silvia R. Vergilio} is currently a professor of Software Engineering in the Computer Science Department of Federal University of Paraná (UFPR), Brazil, where she leads the Research Group on Software Engineering. Her research interests include software testing, software reliability, Software Product Lines (SPLs) and Search Based Software Engineering (SBSE). She serves on the program committee of diverse conferences related to SBSE and software testing, acts as peer reviewer for diverse international journals and is assistant editor of the Journal of Software Engineering: Research and Development.
\end{IEEEbiography}

\balance

\end{document}